\documentclass[twocolumn,prd,floatfix,preprintnumbers,a4paper,nofootinbib,superscriptaddress,groupedaddress]{revtex4-1}
\usepackage[english]{babel}
\usepackage[autostyle, english = american]{csquotes}
\MakeOuterQuote{"}
\usepackage[utf8]{inputenc}
\usepackage{amsmath,amssymb}
\usepackage{amsfonts}
\usepackage{bm}
\usepackage{graphics}
\usepackage{float}
\usepackage{amsmath}
\usepackage{amssymb}
\usepackage[normalem]{ulem} 
\usepackage[mathscr]{euscript}
\usepackage{acronym}
\usepackage{float}
\usepackage[caption=false]{subfig}
\usepackage{lipsum}
\usepackage{mathrsfs}
\usepackage{mathtools}
\usepackage{multirow}
\usepackage{graphicx}
\usepackage{mathtools}
\usepackage{multirow}
\usepackage{graphicx}
\usepackage[usenames,dvipsnames,table,xcdraw]{xcolor}
\usepackage[colorlinks, urlcolor=blue,citecolor=blue,pdfborder={0 0 0}]{hyperref} 
\usepackage[nameinlink,capitalise]{cleveref}
\usepackage{xspace}
\usepackage{xfp}
\usepackage{array}

\newcolumntype{C}[1]{>{\centering\arraybackslash}p{#1}}

\definecolor{veryperi}{RGB}{102, 103, 171}
\definecolor{mygreen}{rgb}{0., 0.5, 0.}

\graphicspath{{fig}}

\graphicspath{{fig}}

\newcommand{\nmax}{N_{\mathrm{max}}}

\begin{document}

\title{Linear vs. nonlinear modelling of black hole ringdowns}
\author{Yi Qiu$^{1,2,3,4}$, Xisco Jim\'enez Forteza$^{3,5,6,7,8}$,
Pierre Mourier$^{8,3,5}$}

\affiliation{$^1$Department of Physics, The Pennsylvania State University, University Park PA 16802, USA}

\affiliation{$^2$ School of Physics, Dalian University of Technology, Dalian 116024, China}

\affiliation{$^3$ Max Planck Institute for Gravitational Physics (Albert Einstein Institute), Callinstra{\ss}e 38, 30167 Hannover, Germany}

\affiliation{$^4$ School of Astronomy and Space Science, Nanjing University, Nanjing 210023, China}

\affiliation{$^5$ Leibniz Universit\"at Hannover, 30167 Hannover, Germany}

\affiliation{$^6$ Nikhef, Science Park 105, 1098 XG Amsterdam, The Netherlands}
 
\affiliation{$^7$ Institute for Gravitational and Subatomic Physics (GRASP),
Utrecht University, Princetonplein 1, 3584 CC Utrecht, The Netherlands}

\affiliation{$^8$ Departament de F\'isica, Universitat de les Illes Balears, IAC3-IEEC, E-07122 Palma, Spain}

\begin{abstract}
The ringdown (RD) phase of gravitational waves is of prime interest for testing general relativity (GR). The modelling of the linear quasi-normal modes (QNMs) within the Kerr spectrum --- or with agnostic parameterized deviations to that GR spectrum --- has become ordinary; however, specific attention has recently emerged to calibrate the effects of nonlinear perturbations for the predominant quadrupolar $l=2$, $m=2$ mode. In this paper, we test the performance of a few nonlinear toy models and of the nonlinear inspiral-merger-ringdown (IMR) model IMRPhenomD for faithfully representing the RD regime and we compare them with the results obtained using linear solutions as sums of QNM tones. Using several quasi-circular, non-precessing numerical waveforms, we fit the dominant $l=2$, $m=2$ mode of the strain, and we assess the results in terms of both the Bayes factor and the inferred posterior distributions for the mass and spin of the final black hole (BH). We find that the nonlinear models can be comparable or preferred over the linear QNM-only solutions when the analysis is performed from the peak of the strain, especially at high signal-to-noise ratios consistent with third-generation observatories. Since the calibration of the tones' relative amplitudes and phases in high-overtone models to the progenitor parameters is still missing, or even not achievable, we consider the use of non-linear models to be more pertinent for performing confident tests of general relativity based on the RD regime starting from early times.

\end{abstract}
\maketitle
\section{Introduction}
\label{sec:intro}
Gravitational waves (GW) provide an excellent arena to study gravity on its strongest regime. 
Since the breakthrough observation of the first event GW150914~\cite{abbott:2016blz}, the GW field itself has experienced an unprecedented growth, as a result of the early-on unexpected but nowadays confirmed high number of GW events observed. Currently, the number of confirmed GW events from compact binary mergers has risen significantly, summing up to about $90$ in the last completed observing run 
(O3)~\cite{LIGOScientific:2018mvr,Venumadhav:2019lyq,LIGOScientific:2020ibl,theligoscientificcollaboration2022gwtc21,LIGOScientific:2021djp,Nitz:2021zwj} of the  LIGO-Virgo-KAGRA (LVK) Collaboration~\cite{theligoscientific:2014jea,thevirgo:2014hva,aso:2013eba}. While these observations have already allowed us to set extraordinary constraints on the general theory of gravity (see, \emph{e.g.}, \cite{tgr,ligoscientific:2019fpa,LIGOScientific:2020tif,LIGOScientific:2021sio}), the near-future prospects are even more promising, with about $200$ new events  anticipated by the end of the current LVK O4 run~\cite{KAGRA:2013rdx}.
The number of observed binary black hole (BBH) mergers is dominating the LVK event catalogue. A typical BBH GW event is described by three different regimes: the inspiral, the merger and the RD. For the optimized search and characterization of the signals, the IMR gravitational waveform templates are used. IMR models are calibrated to numerical relativity (NR) solutions and provide us with the most accurate representation of the full waveform. For BBH mergers, current IMR waveform approximants are normally split into three different families: the IMRPhenom~\cite{Pratten:2020ceb}, SEOBNR~\cite{Ossokine:2020kjp}, TEOB~\cite{Nagar:2023zxh} and NRSurrogates (for a detailed description of the models see~\cite{Garcia-Quiros:2020qpx,bohe:2016gbl,cotesta:2018fcv,Varma:2019csw} and the references therein).


The study of the RD regime has drawn some attention in the last recent years. The RD describes the post-merger phase, in which the final, perturbed BH relaxes rapidly towards its stationary Kerr configuration, a phase which is associated with a characteristic late train of radiation~\cite{leaver:1985ax,detweiler:1980gk,kokkotas:1999bd,press1973,teukolsky1}. Linear perturbation theory provides a simple description of this late radiation regime in the form of a countably infinite sum of damped sinusoids. Each damped sinusoid --- or mode ($lmn$) --- is at most described by four parameters, namely its frequency, damping time, amplitude and phase. The family of frequencies and damping times is known as the quasi-normal-mode (QNM) spectrum~\cite{leaver:1985ax,detweiler:1980gk,kokkotas:1999bd} and, by virtue of the BH no-hair theorem, is uniquely determined by the final BH mass and spin~\cite{Hawking:1971vc,Carter:1971zc,Robinson:1975bv}. The set of amplitudes and phases is determined by the progenitor parameters and the initial orbital conditions~\cite{berti:2005gp,JimenezForteza:2020cve,Forteza:2022tgq,Kamaretsos:2011um,Kamaretsos:2012bs,london:2014cma,London:2018gaq}.
The BH no-hair theorem is tested through two common scenarios: namely, by performing an IMR consistency test~\cite{Hughes:2004vw} and through BH spectroscopy~\cite{Dreyer:2003bv}. BH spectroscopy typically targets the independent estimation of the spectrum of at least two separate RD modes --- although GR deviations can be also measured with a single mode if one considers information from the progenitor BHs~\cite{LIGOScientific:2020tif,LIGOScientific:2021sio,Gennari:2023gmx}. So far, several works from different groups dispute the successful/unsuccessful multiple-mode testing of the theorem with the loud-RD events GW150914~\cite{isi2019,Carullo:2019flw,LIGOScientific:2020tif,CalderonBustillo:2020rmh,Isi:2022mhy,Finch:2022ynt,Cotesta:2022pci,MaSunChen2023a,IsiFarr2023,CarulloReply2023,WangEtAl2023} and GW190521~\cite{LIGOScientific:2020iuh,LVK_190521_analysis,LIGOScientific:2020tif,Capano:2021etf,CapanoEtAl2022,Siegel2023}. In coming years, robust testing of the no-hair theorem~\cite{Gossan:2011ha} through BH spectroscopy may be achieved with loud events at LVK design sensitivity (see~\cite{brito:2018rfr,greg1}) and become more precise at the LIGO $A\#$ sensitivity~\cite{Asharp_sensitivity}, reaching even the percent accuracy with third-generation gravitational wave interferometers such as the Einstein Telescope (ET)~\cite{Maggiore:2019uih}, Cosmic Explorer (CE)~\cite{Evans:2021gyd} and LISA~\cite{lisa}, thanks to the expected and promising gain in signal-to-noise ratios (SNRs).

One of the major debates among the BH ringdown community pertains to the suitability of linear perturbation theory for describing the whole RD regime~\cite{bhagwat:2016ntk,Baibhav:2023clw,Bhagwat:2019dtm,giesler2019,Forteza:2021wfq,isi2019,Carullo:2019flw,Finch:2022ynt,CalderonBustillo:2020rmh,Cheung:2022rbm,Takahashi:2023tkb}. On the one hand, some previous works advocate that the results of linear perturbation theory can be applied from the peak of the gravitational strain onwards, which implies that the non-linear effects observed at the merger regime become quickly irrelevant~\cite{giesler2019,okounkova:2020vwu}. This is assessed by fitting to NR data the RD models with a large number of QNM tones --- typically $N=7$ overtones in addition to the fundamental mode --- to obtain an accurate recovery of the BH parameters, while fixing its QNM spectrum to GR. On the other hand, such claims have been disputed by other works by observing that a high instability of high-overtone models can arise due to i) a likely overfitting of the data and ii) neglecting the yet unmodelled non-linear contributions on the dominant $lm = 22$  mode~\cite{Ripley:2020xby,Forteza:2021wfq,Mourier:2020mwa,Baibhav:2023clw}. In particular, even the linear-order contributions arising from the branch cut\footnote{%
The QNMs of Kerr don't form a complete basis even at linear order~\cite{Green:2022htq,London:2023aeo}. The time-domain Green's function might be split into three different terms, namely, the quasinormal mode solution, the branch cut and a high frequency response. In particular, the prompt emission is originated from the branch cut solution and it is expected to be important at times around the peak of the emission (see~\cite{Leaver:1986gd} for a detailed review).%
}, such as the prompt response or the late tail effects~\cite{Leaver:1986gd,Nollert:1999ji,Ansorg:2016ztf}, are neglected by ringdown models solely based on QNMs. Moreover,~\cite{Cheung:2022rbm,Mitman:2022qdl} have found clear evidences of quadratic contributions in higher harmonics of the RD (specifically, quadratic contributions to the $lm = 44$  mode sourced by the first-order $22$-mode perturbations), which provide more accurate and more stable models than the linear models for these modes. Separately, similar conclusions are obtained by studying the shear at the horizon in head-on BH collisions~\cite{Mourier:2020mwa,Khera:2023lnc}. Unfortunately, an analogous but conclusive analysis for the quadrupolar and dominant $22$ mode in quasi-circular mergers is still absent.

In this work we have tested the performance of linear and non-linear RD models, by fitting the post-peak regime of NR waveforms from the SXS and the (associated) Ext-CCE NR catalogues~\cite{Boyle:2019kee}. The SXS NR waveforms are extracted at finite radii and then extrapolated to future null infinity. The Ext-CCE waveforms use the Cauchy characteristic extraction procedure, thus reducing significantly the gauge dependence of the waveforms obtained at null infinity. We consider the following RD models to fit the data: i) and ii) two RD models described by linear perturbation theory with a variable number of tones, with or without degrees of freedom allowing for restricted deviations from the GR QNM spectrum; iii) the RD sector of the non-linear IMRPhenomD model; and iv) a non-linear RD toy model that uses the linear solution but modifies it slightly to add a non-linear qualitative behaviour at early times. Those models are described in Sec.~\ref{sec:ringdown}. In Sec.~\ref{sec:measure} and Sec.~\ref{sec:setup} we introduce the Bayesian framework and other statistical tools used to perform parameter inference and to assess the physical reliability of the models. In Sec.~\ref{sec:results}, we perform Bayesian parameter inference on a set of zero-noise-realization NR signal injections for each of the models described in Sec.~\ref{sec:ringdown}. Finally we conclude about the accuracy and suitability of each model at describing the RD regime in Sec.~\ref{sec:conclusion}. 
\section{RD models}
\label{sec:ringdown}
%
\subsection{QNM overtone models}
\label{sub:overtone_model}
At late enough times, the RD regime can be modelled \emph{via} the Teukolsky equation~\cite{Teukolsky:1972my}, which describes linear perturbations off a Kerr background spacetime, and hence tells us how GWs propagate as $s=-2$ gravitational perturbations. This equation is typically solved by applying outgoing boundary conditions at null infinity and infalling boundary conditions at the BH horizon. The Teukolsky equation then becomes an eigenvalue problem whose solution is the countably infinite set of the complex QNMs of the final (Kerr) BH. In a time evolution, these modes take the form of exponentially damped sinusoids. Their complex frequencies $\omega_{lmn}=w_{lmn}-\iota/\tau_{lmn} $, corresponding to poles of the Green function~\cite{leaver:1985ax,detweiler:1980gk,kokkotas:1999bd}, are solely determined by the remnant BH's mass $M_f$ and spin $a_f$, in the absence of a BH charge. Here, $\mathrm{Re}[\omega_{lmn}]=w_{lmn}$ are the so-called oscillation frequencies and $-\mathrm{Im}[\omega_{lmn}]=1/\tau_{lmn}$  are the damping rates (inverse of the damping times). These modes are labelled by three integers $l$, $m$, and $n$. Here $l=2,3,\dots$ and $m=-l,-l+1, \dots, l-1, l$ denote the two angular indices of the spheroidal harmonics decomposition. The complex strain at future null infinity $h =h_{+}-i\, h_{\times}$ (where $h_+$ and $h_\times$ denote the two polarization components measured in the detector frame) of the gravitational radiation is accordingly written as:
\begin{equation}
\label{eq:rdharmonicsdecomp}
    h(t,\theta,\phi) = \sum_{l,m,n} h_{lmn} (t)\; {}_{-2}\mathcal{Y}_{lmn}(\theta,\phi;a_f) \, ,
\end{equation}
where $_{-2}\mathcal{Y}_{lmn}(\theta,\phi;a_f)$ are the spin-weighted spheroidal harmonics of spin weight $s = -2$, which depend upon the polar angle $\theta$, the azimuthal angle $\phi$, and the final spin $a_f$.
The third index, $n=0,1,2,\dots$, labels the tones, in order of decreasing damping times $\tau_{lmn}$ for any given $(l,m)$ harmonic. This convention sets the $n=0$ (fundamental) mode as the dominant one at late times while the $n \geq 1$ mode (overtones) are shorter-lived. In addition, there are two branches of QNMs for each $(l,m)$ harmonic, respectively with positive and negative $w_{lmn}$ values~\cite{berti:2009kk}. The counter-rotating modes, with $w_{lmn} < 0$, are nevertheless, expected to have very small relative amplitudes in the dominant harmonics sourced by a quasi-circular merger~\cite{Finch:2021iip,JimenezForteza:2020cve,Dhani:2021vac}, and we shall not consider them further in this paper.

Methods to calculate numerical values of QNM frequencies are present in the literature~\cite{berti-webpage,berti:2005ys,berti:2009kk,cook-QNM,Cook:2014cta,Stein:2019mop,Chung:2023zdq,Chung:2023wkd} for various situations, building upon Leaver's continued fraction method~\cite{leaver:1985ax} or spectral decompositions~\cite{Chung:2023zdq,Chung:2023wkd}. In our work, we mainly use the open-source \texttt{qnm} Python package~\cite{Stein:2019mop} to compute the required Kerr QNM frequencies as functions of the remnant's mass and spin. Alternative open-source algorithms to compute the Kerr spectrum are also available in~\cite{my-codeberg,my-github,vitor-webpage,berti-webpage}.

In this work, we focus on the dominant $(l=2, m=2)$ mode of the strain, $h_{22}(t)$, both in terms of the NR data we are considering and of the models used to describe it. Note that the mode we select from NR simulations is in fact the $(l=2, m=2)$ mode in a \emph{spherical} harmonics decomposition, meaning that it includes some contributions from higher spheroidal harmonics (and the associated QNMs): in aligned-spins cases like we consider, there is in particular some mode mixing with the $(l \geq 3, m=2)$ harmonic. Such contributions are expected to be negligible at all times due to the combination of much smaller amplitudes of these higher modes and small mixing coefficients, so that the spherical $(2,2)$ mode considered is a close approximation to the spheroidal one~\cite{giesler2019,cook2020,Dhani:2021vac}. They might nevertheless affect analyses involving high overtones~\cite{Baibhav:2023clw}, as the small contributions from the fundamental modes ($n=0$) of these other harmonics --- like the $(l=m=2, n=0)$ dominant mode --- are much longer-lived than those overtones. Furthermore, such mode mixing contributions may be enhanced in NR waveforms due to super-translation caused by memory effects, which can possibly be solved by fixing the BMS frame~\cite{Mitman:2022kwt}. We shall however leave such considerations for future work, and neglect mode mixing into the $(2,2)$ spherical mode of the NR strain in the present analyses.

The decomposition of $h_{22}(t)$ into the $(l=2, m=2)$ QNMs up to a given number $N \geq 0$ of overtones defines the linear \emph{overtone model} for the RD:
\begin{equation}
\label{eq:rdwaveansatz}
    \mathrm{OM}_N(t) = \sum_{n=0}^N {\mathcal A}_{n} \, e^{- \iota \,  (t-t_r) \, \omega_{22n} }\, ,
\end{equation}
where $t_r$ is a reference time, usually chosen as a sufficiently late point for the system to reside in the linear regime~\cite{Giesler:2019uxc,london:2014cma,Bhagwat:2017tkm,Bhagwat:2019dtm,forteza2020,London:2018gaq,Carullo:2018sfu}, and
${\mathcal A}_{n}=A_{n} \, e^{\iota \varphi_{n}}$ is the complex amplitude of the $(l=2,m=2,n)$ tone at $t = t_r$. These $N+1$ complex amplitudes, plus the final dimensionless mass and spin $m_f$, $a_f$ which parametrize the QNM frequencies $\omega_{22n}(m_f,a_f)$, define the $2 N + 4$ real-valued free parameters of the model. Refs.~\cite{JimenezForteza:2020cve,london:2014cma} provide separate models of the overtone amplitudes and phases, calibrated to the progenitor parameters limited to $N=1$, while an analogous calibration for $N>1$  is still missing\footnote{%
Ref.~\cite{CalderonBustillo:2020rmh} observes, by performing full Bayesian inference on the event GW150914, that the overtone model $\mathrm{OM}_1$ with the phase $\phi_{221}$ fixed to the NR calibrated value is disfavoured in terms of the Bayes factor with respect to the $\text{OM}_1$ model in which $A_{221}, \phi_{221}$ are freely sampled -- or not constrained to the NR fit values. This discrepancy might be sourced by the strong dependence of the value of the free phase $\phi_{221}$ on the ringdown starting time, as is remarked in~\cite{Forteza:2022tgq}}.
%

This model provides \emph{a priori} a good description of the RD once sufficiently into the linear regime, and while asymptotic late-time non-QNM linear tail contributions are still negligible\footnote{%
Such a tail contribution, with a non-oscillatory power-law decay, is indeed expected as an additional solution to the Teukolsky equation since QNMs do not form a complete solution basis~\cite{Leaver:1986gd,price:1972pw,Nollert:1999ji,Ansorg:2016ztf,London:2023aeo}. Its amplitude is nevertheless small enough that this term only appears at very late times and is typically not visible within the post-merger time range of quasi-circular NR simulations such as the ones we consider~\cite{Leaver:1986gd,Baibhav:2023clw}. Such tail effects have only been observed very recently in eccentric mergers from NR, where they were expected to be enhanced compared to quasi-circular cases~\cite{Carullo2023}.%
}. This model has been argued to be potentially applicable early on in the RD, up to the peak of the strain's amplitude~\cite{giesler2019}, for a sufficiently large $N$. We shall however also consider several alternative models including nonlinear terms, which may better capture the behaviour of the strain close to merger. 
\color{black}  
\subsection{Phenomenological nonlinear toy models}
 The first model with deviations to the linearized GR sum of QNMs that we consider, is based on the parameterized QNM models popular in spectroscopic studies, which allow for deviations of the QNMs to the Kerr spectrum. To avoid the very large parameter space and reduce the possible overfittings of models with multiple overtones that can each independently deviate from GR, we consider the restricted case where only the highest tone of the model is modified. This defines the `highest-tone perturbation models' HTPM:
\begin{align}
\label{eq:HTPM}
   & \mathrm{HTPM}_N(t) \equiv \mathrm{OM}_{N-1}(t) + {} \nonumber \\
   & \qquad {\mathcal A}_{N} \, e^{- \iota \,  (t-t_r) \, w_{22N} (1+\alpha_N)} \, e^{- (t-t_r) / (\tau_{22N} (1+\beta_N))} \; ,
\end{align}
where $\alpha_N$ and $\beta_N$ are respectively the oscillation frequency and damping time perturbation parameters. Their values measure the deviations of the highest included tone from the Kerr spectrum, and the QNM solution from GR would be recovered for $\alpha_N=\beta_N=0$. $\mathrm{HTPM}_1$ corresponds to the most widely used QNM model in overtone-based spectroscopy (\emph{e.g.} \cite{isi2019,CalderonBustillo:2020rmh,LIGOScientific:2020tif,LIGOScientific:2021sio,MaSunChen2023a}), including the $220$ and $221$ modes with possible frequency and damping times deviations to the Kerr values for the first overtone. $\mathrm{HTPM}_{\geq 2}$ is a restricted extension of this model to higher numbers of overtones, where only the last tone is allowed to deviate from the GR spectrum. $\mathrm{HTPM}_0$ is also formally well-defined by the above Eq.~\eqref{eq:HTPM}, but it has very limited interest due to the full degeneracy between the free final mass and spin and the deviation parameters on the complex frequency of the only mode present ($n=0$), making it effectively an over-parametrized version of $\mathrm{OM}_0$. We will accordingly not consider it in our analyses. Note that $\mathrm{HTPM}_N$ depends in total on $2 N + 6$ real-valued free parameters (\emph{i.e.}, the same number as $\mathrm{OM}_{N+1}$): the final mass and spin $(M_f, a_f)$; the perturbation parameters $\alpha_N$ and $\beta_N$; and the complex amplitudes of the $N+1$ tones included.

The other phenomenological toy model considered in this work amounts to a sum of QNMs with a nonlinear transformation of the time coordinate. The time parameter is shifted by an exponentially decaying term, so that the model can exhibit this nonlinear modification at early times close to merger (where it leads to a slower variation of the phase and amplitude of the waveform), while asymptotically recovering the linear model at later times\footnote{Since the prompt response is expected to be higher close to the peak of the emission, the TCTM model is designed to account for this excess, although its fundamental form remains unknown. The TCTM may as well be interpreted as phenomenologically accounting for post-merger GWs from a BH of time-dependent mass and spin due to the absorption of the infalling radiation~\cite{Papadopoulos:2000gb,Sberna:2021eui,Redondo-Yuste:2023ipg}.} We define the corresponding `time coordinate transformation models' TCTM as follows:
\begin{equation}
    \label{eq:TCTM}
 \mathrm{TCTM}_N\left(t\right) \equiv \mathrm{OM}_N\left(t+A e^{-t/\tau}\right) \; ,
\end{equation}
where $A$ and $\tau$ are two constant parameters to be determined. Like in the HTPM case above, $\mathrm{TCTM}_N$ depends on $2 N + 6$ real-valued free parameters: the final mass and spin; $A$; $\tau$; and the $N+1$ complex tone amplitudes.

We also considered several additional phenomenological toy models, introduced in~\cite{Dhruv} for the description of the decaying deformations of the final horizon formed in a merger. These models corresponds to overtone models modified \emph{e.g.} by the addition of a non-oscillatory exponential decaying component (under the form $\mathrm{OM}_N(t) + B \exp [-(t-t_r) / \tilde\tau]$), or of a power-law decaying term with or without oscillations ($\mathrm{OM}_N(t) + C (t-t_1)^{-\gamma} \exp[\iota \, \omega (t-t_r) ]$, or the same form with $\omega$ set to $0$). These models are designed to recover the steep decay featured by the deformations of the common horizon shortly after its formation, in addition to the late-time QNM oscillations. Hence, these models were not expected to be necessarily also well-suited to the description of ringdown waveforms at null infinity, which rather have a slower change of amplitude near the merger than further into the RD regime. Indeed, the preliminary comparative tests of the various models discussed in this section (see below in Sec.~\ref{subsec:prelim_tests}) indicated a poor performance for these additional models for the description of waveform ringdowns, and we did not consider them further for the present work.

\subsection{IMR model}
\label{sub:phenomenological-model}
One can also use full inspiral-merger-ringdown waveform models to describe the GW ringdown, by selecting only the post-merger part of these models. They are built upon non-linear ans\"atze, calibrated to NR waveforms, which depend solely on the progenitors' parameters. In general, they will require fewer input parameters than the overtone models --- especially compared to those including a large number of overtones $N_{\max}$. In this work, we use the nonprecessing "IMRPhenomD" waveform model. This model is calibrated to mass ratios up to $q=1/18$ and initial effective spins up to $0.98$~\cite{Khan:2015jqa}, and is well sufficient to cover the non-precessing, $22$ mode--only waveforms we consider in this work. Another advantage of using IMRPhenomD for this study is that it has been calibrated to a relatively small number of numerical waveforms, allowing us to easily pick a set of NR waveforms for the present study that were not involved in that calibration\footnote{%
We have also tested two other waveform approximants for comparison, namely IMRPhenomPv2 and SEOBNRv4, in the case of analyses starting at the strain amplitude peak ($t_0 = 0$), and found similar or better performances than IMRPhenomD. We chose IMRPhenomD in part because of the fact that more state-of-the-art approximants were usually calibrated to a much larger number of NR waveforms~\cite{bohe:2016gbl,Pratten:2020ceb,PhenomXO4a}, and thus potentially exhibit artificial advantages in the model comparison.}%
.

We generate the IMRPhenomD using the PyCBC~\cite{alex_nitz_2023_8190155} time-domain approximant. We truncate the waveform at $t = 0$, aligning it with the peak of the strain for the 22 mode. This alignment corresponds to a frequency $f_{min} \sim 170$ Hz. We use a sampling rate $\delta t=1/2048$ s, appropriate to resolve the high-frequency RD regime. We select only the $22$ mode. The inclination angle is fixed to zero so that the complex strain is just $h_+ - i h_\times = h_{22} (t) {}_{-2}\mathcal{Y}_{22}(\theta,\phi)$. Note that we are comparing the IMRPhenomD model with NR waveforms, hence, we must translate the IMR output to geometric units with $G =c=M=1$, which implies that the final mass referred in this work represents the fraction of energy radiated $M_f/M = 1$.  With this setup, we are left with 4 free parameters in the model: mass ratio, $\chi_1^z$, $\chi_2^z$, and phase, where $\chi_1^z, \chi_2^z$ denote the $z$ component of the initial dimensionless spin of the progenitor BHs. Finally, the final mass $M_f$ and final spin $a_f$ are obtained from fits to NR in terms of the progenitor parameters~\cite{jimenez-forteza:2016oae,hofmann:2016yih}.

\section{Model comparison}
\label{sec:measure}

\subsection{Mismatch analyses}
\label{sec:mis}
While the bulk of our analysis is based on a full Monte-Carlo sampling parameter estimation for each model, we first performed a least-mean-squares fit-based preliminary comparison and selection among the models at hand. The modelling quality can be assessed in this case using the mismatch $\mathcal{M}$. This quantity commonly used in GW astronomy to quantify the similarity between the model and data~\cite{Giesler:2019uxc,Bhagwat:2019dtm,forteza2020} is defined as:
\begin{equation}
    \mathcal{M} = 1 - \frac{\langle h_{\rm NR}|h_m\rangle}{\sqrt{\langle h_{\rm NR}|h_{\rm NR}\rangle \langle h_m|h_m\rangle}}\,
    \label{eq:mismatch}
\end{equation}
where we compare between our model for the strain $22$ mode, $h_m(t)$, and the corresponding numerical data, $h_{\rm NR}(t)$, with:
\begin{equation}
\label{eq:inner}
    \langle f|g\rangle = \mathrm{Re} \int_{t_0}^{t_f} f^*(t) g(t) \, \mathrm{d}t \; .
\end{equation}
The limits of the integral $t_0$ and $t_f$ mark the fit starting and ending time, respectively. $\mathcal{M}$ varies between 0 and 1, with higher mismatches meaning that the model deviates more from data. We can thus compare models by computing the mismatch of each of them (with the best-fit parameter values) to the data. A lower value of that mismatch is associated with a more faithful modelling --- without, at this stage, accounting for possibly different numbers of free parameters.

\subsection{Bayesian framework}
Apart from mismatches, another quantity of common use for comparing the fitting performance of different GW models is the Bayes factor. As opposed to the mismatch, the Bayes factor directly quantifies the relative evidence in favour of a model versus another given the data, and this comparison already includes an implicit penalty cost for additional parameters and potential overfitting, through the prior and parameter space volume. It applies within a Bayesian parameter inference framework for each model $\mathscr{M}$, where the posterior probability distribution of the parameters $\vec{\theta}$ associated with the model, given the data $d$ (here $d= h_{\mathrm{NR}}$), is given by:
\begin{equation}
    \label{eq:bayes}
    p(\vec{\theta} \, | \, d, \mathscr{M})= \frac{p(\vec{\theta} \, | \, \mathscr{M}) \, p(d \, | \, \vec{\theta}, \mathscr{M})}{p(d \, | \, \mathscr{M})} \; .
\end{equation}
Above, $p(\vec{\theta}\,  | \, \mathscr{M})$ is the prior probability of the parameters $\vec{\theta}$ , $p(d \, |\, \vec{\theta}, \mathscr{M})$ is the likelihood function which represents the conditional probability of observing $d$ given the model $\mathscr{M}$ with parameters $\vec{\theta}$, and $\mathcal Z \equiv p(d \, | \, \mathscr{M}) = \int p(d \, | \, \vec{\theta}, \mathscr{M}) \, p(\vec{\theta} \, | \, \mathscr{M}) \, \mathrm{d} \vec\theta$ is the evidence associated with model $\mathscr{M}$. 

Two possible models $\mathscr{M}_A$ and $\mathscr{M}_B$ for the description of a given dataset $d$ can then be compared by computing their relative Bayes factor, \emph{i.e.}, the ratio of evidence between them:
\begin{equation}
    \label{eq:bayes factor}
    \mathcal{B}_{A B}=\frac{p\left(d \mid \mathscr{M}_{A}\right)}{p\left(d \mid \mathscr{M}_{B}\right)} \; .
\end{equation}
As per usual practice, and for convenience, we will be using the logarithmic Bayes factor $\log_{10} \mathcal{B}_{A B}$, as well as the logarithmic evidence for any given model $\log_{10}\mathcal{Z}$ (the difference of logarithmic evidences between two models providing directly the logarithmic Bayes factor between them). Note that we use the base-$10$ log here.
$\mathcal{B}_{A B} > 1$ ($\log_{10}\mathcal{B}_{A B} > 0$) would support model $\mathscr{M}_A$ over model $\mathscr{M}_B$, and vice versa; although a significant claim of preference of $\mathscr{M}_A$ over $\mathscr{M}_B$ would require $\log_{10}\mathcal{B}_{A B} \gtrsim 1$~\cite{Kass:1995loi}. 

\subsection{Bias analyses}
\label{sec:eps}
Typically, both the best-fit mismatch and Bayes factor (with respect to a reference model) for each model are only used to assess the fitting quality, without accounting for the physical accuracy of the values obtained for the model parameters. This accuracy can be measured separately \emph{via} the combined recovery bias $\epsilon$ on the final mass $M_f$ and dimensionless final spin $a_f$~\cite{Giesler:2019uxc,Forteza:2021wfq}:
\begin{equation}
    \label{eq:epsilon}
    \epsilon = \sqrt{\left(\frac{M_f^{\mathrm{fit}}-M_f^{\mathrm{true}}}{M}\right)^2 + \left(a_f^{\mathrm{fit}}-a_f^{\mathrm{true}} \right)^2} \; ,
\end{equation}
where the final masses are normalized by the initial total mass $M$ from the NR simulation. The true parameters $M_f^{\mathrm{true}}$ and $a_f^{\mathrm{true}}$ correspond to the values from the NR simulation, that are estimated from the mass and spin quasi-local definitions on the apparent horizon~\cite{Boyle:2019kee,Iozzo:2021vnq,Forteza:2021wfq,Ashtekar:2004cn}. The uncertainties of these final mass and spin local estimates from NR are typically in the order of $10^{-5}$ to $10^{-4}$~\cite{Forteza:2021wfq}.

\begin{figure*}
\includegraphics[width=2\columnwidth]{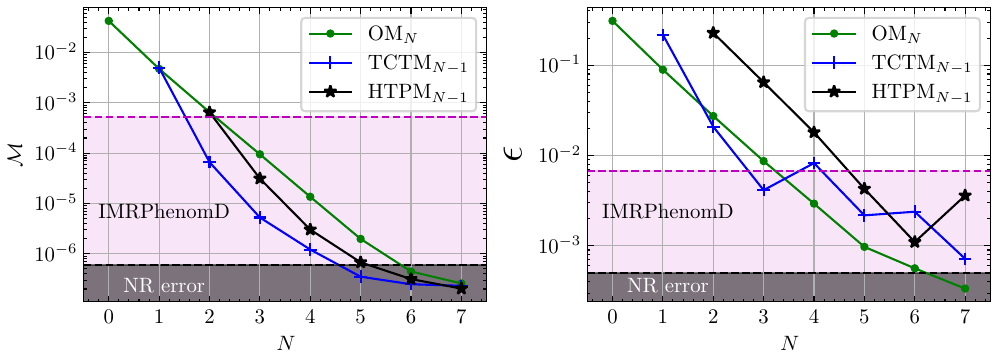}
\caption{In the left panel, we show the evolution with $N \in \{0, \dots, 7 \}$ of the mismatch $\mathcal{M}$ between the NR waveform and the best-fit $\mathrm{OM}_N$, $\mathrm{TCTM}_{N-1}$ and $\mathrm{HTPM}_{N-1}$ models (discarding the over-parametrized model $\mathrm{HTPM}_0$). For a given $N$ (along vertical lines on the figure), these models have the same number ($2 N +4$) of free parameters and can thus be directly compared. The NR ringdown waveform considered here corresponds to the BBH:0305 simulation from the SXS catalog, selecting the $22$ mode and starting at its amplitude peak ($t=0$). In the right panel, the evolution with $N \in \{0, \dots, 7 \}$ of the bias parameter $\epsilon$ is presented for the same models. The magenta dashed line and shaded area in each panel mark the optimal mismatch and bias obtained for the IMRPhenomD model, which has $4$ free parameters like the $\mathrm{OM}_0$ ($N=0$) model. The dark grey dashed line and shaded area at the bottom of each panel delimit the mismatch and $\epsilon$ values that are lower than the estimated NR errors, \emph{i.e.}, the maximum of the mismatches $\max \left(\mathcal{M}_{\text {res }}, \mathcal{M}_{\text {extr }}\right)$ and the total discrepancy $\delta \epsilon_{\mathrm{r}}$ on the dimensionless radiated energy and angular momentum, obtained from comparing the two highest available resolutions and from comparing the two best extrapolation orders for the $22$ mode of the NR waveform (see Sect.III-B of~\cite{Forteza:2021wfq}). Note that the minimum grid resolution used for the recovery of $M_f/M$ and $a_f$ for the OM, HTPM and TCTM models was $| \delta M_f|/M, |\delta a_f | = 3.2 \cdot 10^{-6}$ for both parameters, and hence has a negligible impact on the $\epsilon$ values obtained.}
\label{fig:modcom}
\end{figure*}

\subsection{Preliminary tests}
\label{subsec:prelim_tests}
As a preliminary step to rapidly select relevant models and assess general trends, we fitted various models, with a range of numbers of tones, to the $h_{22}$ ringdown strain mode from the BBH:0305 binary BH simulation from the SXS catalogue~\cite{Boyle:2019kee}. The ringdown phase of the signal was selected by starting the analysis at the peak of the corresponding amplitude $|h_{22}|$ (which we use to define $t=0$), \emph{i.e.}, by setting $t_0 = 0$; up to the end of the available data ($t_f / M \sim 150$). For the sake of stability and efficiency of the fits, we followed the same algorithm as in~\cite{Forteza:2021wfq,my-codeberg} to obtain the best-fit parameters for each model. The two to four parameters of a given model which are not tone amplitudes and phases (that is, for instance, $\{M_f, a_f \}$ for overtones models, and $\{ M_f, a_f, A, \tau \}$ for the HTPM models) are distributed on an adaptative grid. For each value of those parameters on the grid, the best-fit value for the remaining parameters of the model, \emph{i.e.} the tone complex amplitudes, on which the model depends linearly, is obtained through the analytic minimization of the sum of squared residuals for a linear model. The mismatch between the corresponding best-fit waveform and the NR waveform is computed for each grid point, and the optimal value for the `nonlinear' parameters on the grid such as $M_f$ and $a_f$ is then determined as the value minimizing the mismatch. Note that minimizing over the mismatch or over the sum of squared residuals, for any given set of parameters, gives equivalent results in the limit of small mismatches (or residuals).

In Fig.~\ref{fig:modcom}, we show the mismatch $\mathcal{M}$ (left panel) and final mass/spin bias $\epsilon$ (right panel) of these best-fit solutions for the OM and the alternative TCTM and HTPM models, at various numbers of overtones. For each value of $N = 1 \dots 7$ on the x axis, we show the results for the $\mathrm{OM}_N$, $\mathrm{TCTM}_{N-1}$ and $\mathrm{HTPM}_{N-1}$ models (plus $\mathrm{OM}_0$ for $N=0$), for direct comparison of models which have the same number of free parameters, \emph{i.e.} $2 N + 4$. We also considered the IMRPhenomD model, which has 4 parameters like the $\mathrm{OM}_0$; we represent its mismatch and bias as a reference magenta horizontal dashed line and shaded area on each panel. The estimated numerical error on the mismatch and bias parameter are also shown as dark gray horizontal lines and shaded areas.

These comparison plots show that the TCTM (at any $N$) and the HTPM (from $\mathrm{HTPM}_1$ onwards) provide a comparable or (in most cases) better description in terms of mismatch to the ringdown $22$ mode of SXS:BBH:0305, than the OMs for the same number of free parameters. On the other hand, these two classes of models typically recover more poorly the parameters of the final BH than the corresponding overtone model, with the notable exception of $\mathrm{TCTM}_1$ and $\mathrm{TCTM}_2$ (compared to $\mathrm{OM}_{2,3}$); while like for the overtone models the bias does decrease in most cases when more tones are included in the model. Finally, the $4$-parameter IMRPhenom model provides a description of the NR data of a quality comparable to the 8-parameter $\mathrm{OM}_2$ and $\mathrm{HTPM}_1$ models (and better than the 6-parameter $\mathrm{TCTM}_0$), with a much lower bias on the final mass and spin. For this model, a good fitting quality could be expected to some extent, thanks to its prior calibration to other NR waveforms including times around the strain peak ($t=0$). Notice, however, that the calibration of IMR approximants is performed and tested as a full inspiral-merger-ringdown model, with its accuracy tested as a whole too. Therefore, the accuracy and performance of the ringdown part alone, is not fully guaranteed \emph{a priori}, especially the good recovery of the final black hole parameters that we find here.

The few other phenomenological RD models, not shown here, that we considered in this preliminary study (\emph{e.g.} adding a non-oscillatory damped power-law or exponential term to a sum of QNM overtones) were typically performing much more poorly in both mismatch and bias than the models shown, and are consequently not considered further in this work.

We shall now focus on the models with low or moderate numbers of parameters which showed comparable or better performance than the first few overtone models, \emph{i.e.}, $\mathrm{OM}_{0 \dots 4}$, $\mathrm{TCTM}_{0 \dots 2}$, $\mathrm{HTPM}_{1,2}$, and IMRPhenomD, for a more thorough comparison in a Bayesian inference approach using nested sampling~\cite{Skilling:2006gxv}.

\section{Parameter estimation}
\label{sec:setup}
For GW detectors, the frequency-domain likelihood function under stationary Gaussian noise is defined from the noise-weighted frequency-domain inner product $(\cdot,\cdot)$ as follows~\cite{Maggio:2022hre}:
\begin{equation}
    \label{eq:likelihood1}
    p(d \,|\, \vec{\theta}, \mathscr{M}) = \exp \left[- \frac{1}{2} \big( d(f)-m(\vec{\theta},f), d(f)-m(\vec{\theta},f)\big) \right] \, ,
\end{equation}
where $m(\vec{\theta},f)$ denotes the strain $22$ mode from model $\mathscr{M}$ with particular parameters $\vec{\theta}$, evaluated at frequency $f$, and $d$ is the sum of the GW strain and the noise realization~\cite{Cabero:2019zyt} for the $22$ mode. The inner product $(\cdot,\cdot)$ itself is defined from the one-sided power spectral density $S_{n}(f)$ of the detector's noise, as:
\begin{equation}
\label{eq:innerFreq}
(x , y) = 4\times \text{Re} \int_{0}^{\infty} \frac{x^{*}(f) \, y(f)}{S_{n}(f)} \, \mathrm{d} f \; .
\end{equation}
In our case, the parameter estimation is performed on time-domain data, with the NR ringdown $22$-mode strain as our data set $d(t)$, considered as being injected in a zero-noise realization. We moreover consider here a flat noise spectrum within the relevant frequency range, \emph{i.e.}, that the noise sensitivity curve can be approximated by its value at the frequency of the $(220)$ mode~\cite{berti:2005ys}, $S_n(f)\approx S_n(f_{220}) =cst$. This assumption allows us to directly relate this constant noise amplitude to the optimal  $\rho$ of the data, with $\rho^2 = (d(f),d(f)) = 2 \, \langle d(t) \,|\, d(t) \rangle / S_n(f_{220})$, and to convert more generally \emph{via} Parseval's theorem the noise-weighted frequency-domain inner product $(\cdot,\cdot)$ of Eq.~\eqref{eq:innerFreq} into the time-domain scalar product of Eq.~\eqref{eq:inner}:
\begin{equation}
    (x(f),y(f)) = 2 \, \frac{\langle x(t) \,|\, y(t) \rangle}{S_n(f_{220})} = \rho^2 \, \frac{\langle x(t) \,|\, y(t) \rangle}{\langle d(t) \,|\, d(t) \rangle} \; .
\end{equation}
The likelihood expression, Eq.~\eqref{eq:likelihood1}, can thus be re-expressed in time domain as:
\begin{equation}
    \label{eq:likelihood2}
    p(d \,|\, \vec{\theta}, \mathscr{M}) = \exp \left[-\rho^2 \:\! \frac{\langle d(t)-m(\vec{\theta},t) \,|\, d(t)-m(\vec{\theta},t)\rangle}{2\,\langle d(t) \,|\, d(t)\rangle} \right]  .
\end{equation}
We can therefore directly set the optimal  $\rho$ in our parameter estimations through the likelihood function, which for a given data $d$ is equivalent to setting the constant noise amplitude within this approximation.

To perform parameter estimation and both obtain posterior distributions and calculate the models' Bayesian evidences, we use the dynamical nested sampling method from the \textit{dynesty} Python package~\cite{Speagle:2019ivv}. One particular feature of this method is that it estimates the evidence and the posterior simultaneously.
Throughout our tests, we used $2000$ n-live points and a stopping criterion of $\Delta (\ln \mathcal{Z})=0.1$ for the nested samplings.

\section{Results}
\label{sec:results}
We start our initial inference analysis on the GW150914-like simulation, SXS:BBH:0305, employing the parameter estimation framework this waveform into white Gaussian noise, simulating an event observed by third-generation (3G) observatories with a signal-to-noise ratio of $\rho = 100$. SXS:BBH:0305 is consistent with a signal with masssratio $q=m_{1}/m_{2}=1.22$, effective dimensionless spin $\chi_\mathrm{eff}=\left(\chi_{1} m_{1}+\chi_{2} m_{2}\right) /\left(m_{1}+m_{2}\right)=0.01$. Our Bayesian inference is performed using four RD models described in Section~\ref{sec:ringdown}. Further details on the various prior choices for each RD model and their impact on the obtained posterior distributions are provided in Section~\ref{app:prior}.
%

\subsection{Bayes factor analysis for $t_0=0$}
\label{sub:results_bayes}
We initiate our parameter estimation (PE) analysis at $t_0 = 0$, i.e., consistent with the peak of the strain. We first estimate the values of the evidence $\mathcal{Z}$ for each of the models considered here. The results of these runs are shown in Table~\ref{table:0305}. We can deduce from the results that the IMRPhenomD model yields the highest value for $\log_{10} \mathcal{Z}$, suggesting a better fit to the NR data compared to the other models. Moreover, among the non-linear models examined in this study, the second and third highest-ranking models in log evidence are the non-linear TCTM models. Specifically, the $\text{TCTM}_2$ yields a log Bayes value of $\log_{10}\mathcal{B}\sim 3$, indicating that it is approximately $\mathcal{O}(10^3)$ times larger than the log evidence for the overtone solutions $\text{OM}_{2}$ and $\text{OM}_{3}$. Notably, the $\text{TCTM}_2$ achieves this superior performance while having the same number of parameters as the latter. We also observe that, among the linear solutions, the $\log_{10} {\mathcal{Z}}$ saturates at $\text{OM}_{2}$ and it provides a slightly lower value for $\text{OM}_{3}$. We verified that the $\log_{10}\mathcal{Z}$ value consistently decreases for $\text{OM}_4$ and $\text{OM}_5$ thus, showing that the optimal performance of the OM models occurs at $\text{OM}_{2}$. This is compatible with the overtone-models results of~\cite{Baibhav:2023clw} and --- in the complementary context of deformations of the final horizon --- of~\cite{Mourier:2020mwa}. Finally, the non-GR model $\text{HTPM}_1$ provides similar $\log_{10} \mathcal{Z}$ as $\text{OM}_2$.

To ensure the robustness of these results and rule out potential influences from varying prior choices, we have examined and confirmed that adjusting priors does not qualitatively alter these findings. The results shown for different prior choices is shown in Appendix~\ref{app:prior}. Therefore, and from the Bayes factors' point of view, the nonlinear models are preferred for the waveform SXS:BBH:0305.

In the fourth column of the same table, we show the mass/spin recovery bias $\epsilon$ as defined in Eq.~\eqref{eq:epsilon} for the ten models considered. The true values $M_f^\mathrm{true},a_f^\mathrm{true}$ are obtained from the NR metadata files, while the $M_f^\mathrm{fit},a_f^\mathrm{fit}$ correspond to the maximum likelihood values obtained after sampling the likelihood distribution of Eq.~\eqref{eq:likelihood2}. The biases on the final mass and the final spin for the OM models improve gradually from $\text{OM}_0$ to $\text{OM}_3$, and it starts to degrade for $\text{OM}_4$ (consistently with the overtone-models analyses of~\cite{Baibhav:2023clw}). Notice that this last result slightly differs qualitatively from the analysis we have done based on the value of the evidence $\mathcal{Z}$. This is expected since the computation of the evidence provides an extra penalty factor to the increase of the prior volume, which in this case is sourced by the large number of free parameters of the high-$N$ OM models. Moreover, we find some mild disagreements between the values of $\epsilon$ obtained in Fig.~\ref{fig:modcom}, with respect to the ones listed in Table~\ref{table:0305}, especially for the models with a high number of overtones, leading to a different behaviour for $\mathrm{OM}_4$. Fig.~\ref{fig:modcom} is obtained by non-linear minimization to fit the NR data, with the precision of $\epsilon$ fixed by the mass-spin grid resolution. Therefore, data overfitting/underfitting may happen with such prescription. On the contrary, the results shown in Table~\ref{table:0305} are obtained from sampling the whole ($2N+4$)-dimensional likelihood distribution, which involves a larger exploration of the parameter space. This reduces the risk of overfitting the data and thus provides a more statistically robust representation of the recovery biases. The statistical uncertainties of $\epsilon$, at the SNR we consider here ($\rho = 100$), will be better illustrated in the next subsection.

Regarding the non-linear models, we obtain a slightly larger bias for the nonlinear IMRPhenomD model than for $\mathrm{OM}_3$, $\epsilon^{\text{IMRPhenomD}} \ \gtrsim \  \epsilon^{\text{OM}_3}$, while the evidence $\mathcal{Z}$ for the IMRPhenomD is much larger. The higher Bayesian evidence for this particular model is due in part to its calibration to a set of NR waveforms (which does not include SXS:BBH:0305), which allows it to depend only on the few parameters characterizing the progenitor BHs: mass ratio, magnitudes of both aligned spins, orbital phase. It was not, however, separately calibrated to the ringdown phase, and the still rather good recovery of the final BH parameters from the whole ringdown (starting at the $t=0$ amplitude peak) that we obtain was not fully ensured by design. The best model in terms of $(M_f, a_f)$ bias is other non linear model , $\text{TCTM}_2$, with $\epsilon=0.0045$.
\begin{table}[h]
\begin{tabular}{C{0.11\textwidth}C{0.11\textwidth}C{0.11\textwidth}C{0.11\textwidth}}
\hline\hline
Model & Parameters & $\log_{10}\mathcal{Z}$ & $\epsilon$\\ \hline\hline
$\text{OM}_0$ & 4    & -188.435  & 0.311261 \\
$\text{OM}_1$ & 6    & -31.9132  & 0.087670 \\
$\text{OM}_2$ & 8    & -17.0332  & 0.027019 \\
$\text{OM}_3$ & 10   & -17.1311  & 0.011475 \\
$\text{OM}_4$ & 12   & -17.7541  & 0.032821 \\
$\text{TCTM}_0$ & 6    & -33.2123  & 0.216384 \\
$\text{TCTM}_1$ & 8    & -14.4255  & 0.024559 \\
$\text{TCTM}_2$ & 10   & -14.2678  & 0.004472 \\
$\text{HTPM}_1$ & 8   & -17.1984  & 0.236650 \\
$\text{HTPM}_2$ & 10   & -17.3353  & 0.096609 \\
IMRPhenomD & 4    & -9.57672  & 0.014981 \\
\hline\hline
\end{tabular}
\caption{
The table contains $\log_{10}$ evidences, and remnant properties recovery biases corresponding to the maximum-likelihood values of the samplings, for the 11 models considered here, for the GW150914-like simulation waveform SXS:BBH:0305 from the SXS collaboration. The sampling is performed using the \textit{dynesty} Python package. The number of free parameters of each model is also given in the second column. Noteworthy, we have found that at $\rho = 100$ as we used here, the value of $\epsilon$ is mildly subjected to finite sampling uncertainties.}  
\label{table:0305}
\end{table}

\subsection{Mass and spin posterior distributions for $t_0=0$}
\label{sub:results_posterior}
We study the marginalised posterior distribution of remnant properties for each of the models considered in this work. Notice here that the posterior distribution of IMRPhenomD are originally $q$, $\chi_{1z}$, $\chi_{2z}$, which are parameters of the progenitor BHs. Thus, these are converted to the final state parameters using NR fits to the final BH state~\cite{jimenez-forteza:2016oae,hofmann:2016yih,alex_nitz_2023_8190155}.

\begin{figure}
\includegraphics[width=0.98\columnwidth]{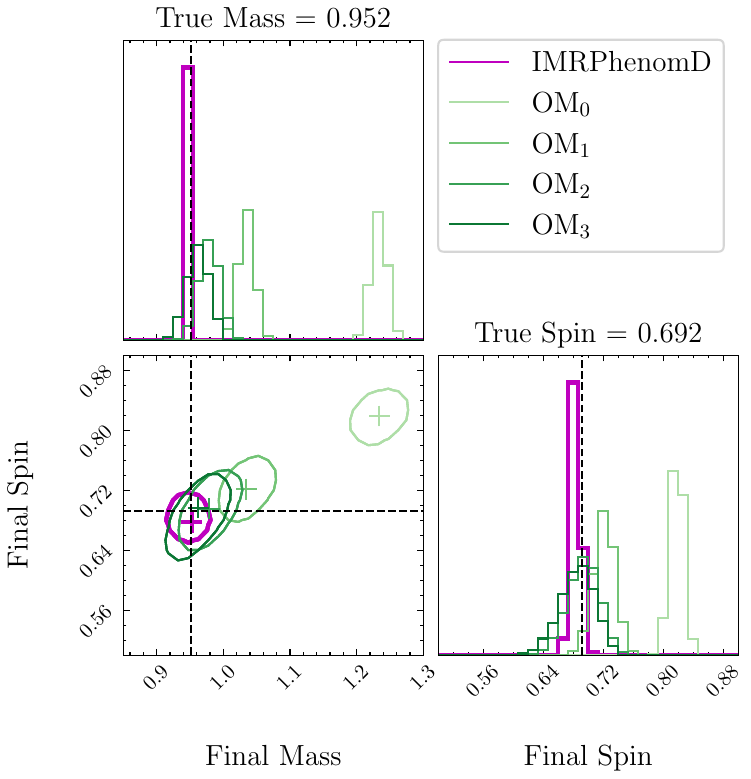}
    \caption{We show the comparison of final mass and spin posterior distribution for the sampling of $\text{OM}_{0-3}$ and IMRPhenomD for waveform SXS:BBH:0305. Each contour represents a $90\%$ credible region on the mass-spin 2D plane for given model. The black dashed lines note the "true" final mass and spin. The "+" signs denote the maximum likelihood values for each of the models considered. We assume the injection SNR $\rho$ to be 100 for all the cases, in order to see the clear comparison. Notice that the 2D marginalised distributions for $\text{OM}_2$ and $\text{OM}_3$ are practically overlapping, thus showing no significant performance gain for the $\text{OM}_3$. Note that we observe a significant overlap between the contours of $\text{OM}_3$ and $\text{OM}_4$. Since we also see in Table.~\ref{table:0305} no gain in Bayesian evidence and recovery bias when including more than 3 overtones, $\text{OM}_4$ is excluded for the clarity of this plot.}
    \label{fig:Mass_Spin_Comparison_OM}
\end{figure}

In Fig.~\ref{fig:Mass_Spin_Comparison_OM}, we show the final mass and final spin posterior distributions for the models $\text{OM}_{0-3}$ and IMRPhenomD (in magenta). The "+" symbols, represent the maximum likelihood values for each of the models considered, and that correspond to the values of $\epsilon$ shown in the fourth column of Table~\ref{table:0305}. Notice that the posterior distributions on the final mass and the final spin are consistent with the true parameters for the models with $\text{OM}_{n \geq 2}$, while it shows large offsets for the models $\text{OM}_{n=0,1}$. Moreover, it is noteworthy that the $90\%$ credible contours increase slightly with the overtone index, from $n=0$ to $n=4$. The broadening of the posterior contours might be sourced by the observed correlations among the tones (see the Appendix B of~\cite{Bhagwat:2019dtm}), which become especially relevant for tones with $n\gtrsim 1$. Therefore, the precision at which one can measure the final mass and final spin for signals at $\rho =100$ using OMs reaches a maximum for $n=2$ and it slightly degrades for $n>2$. Alternatively, notice that the IMRPhenomD model provides i) lower values for the bias $\epsilon$ than the best OM solution, and ii) tighter mass and spin contours than all the other models considered. Therefore, we observe that the IMRPhenomD model provides a more accurate description of the signal from $t_0=0$ than the OM models, both if one uses the Bayesian evidence as a ranking criterion, and from testing the consistency of the posterior distributions with respect to the true parameters. 

\begin{figure}
\includegraphics[width=0.98\columnwidth]{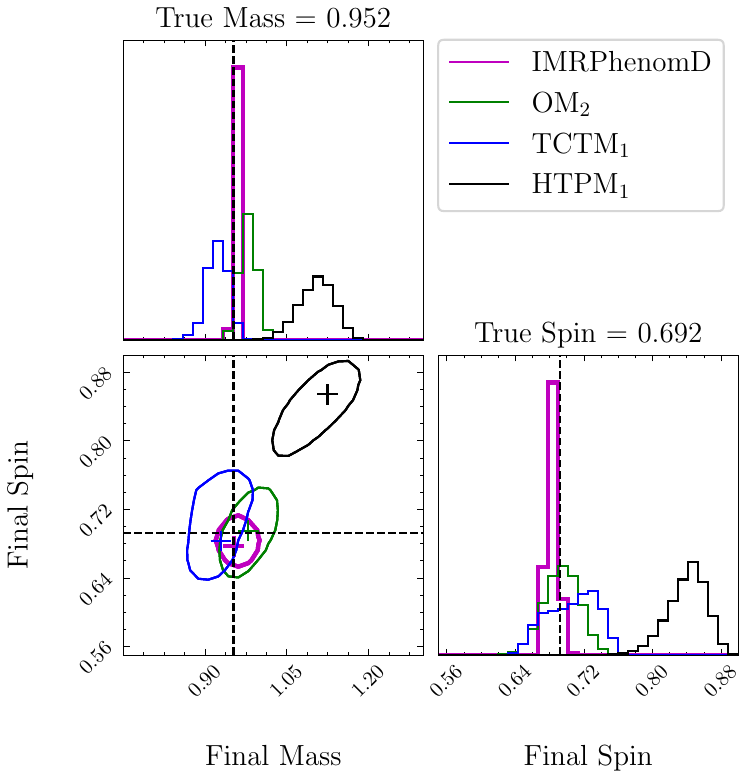}
    \caption{We show the comparison of final mass and spin posterior distribution for the sampling of $\text{OM}_2$, $\text{TCTM}_1$, $\text{HTPM}_1$ and IMRPhenomD for the NR waveform SXS:BBH:0305. Each contour represents a $90\%$ credible region on the mass-spin 2D plane for a given model. The black dashed lines note the "true" final mass and spin. The "+" symbols represent the maximum likelihood estimated for each of the models considered in this work. We assume the injection SNR to be 100 for all the cases, in order to see the clear comparison. 
    }
    \label{fig:Mass_Spin_Comparison_model}
\end{figure}

\begin{figure}[h]
\includegraphics[width=0.98\columnwidth]{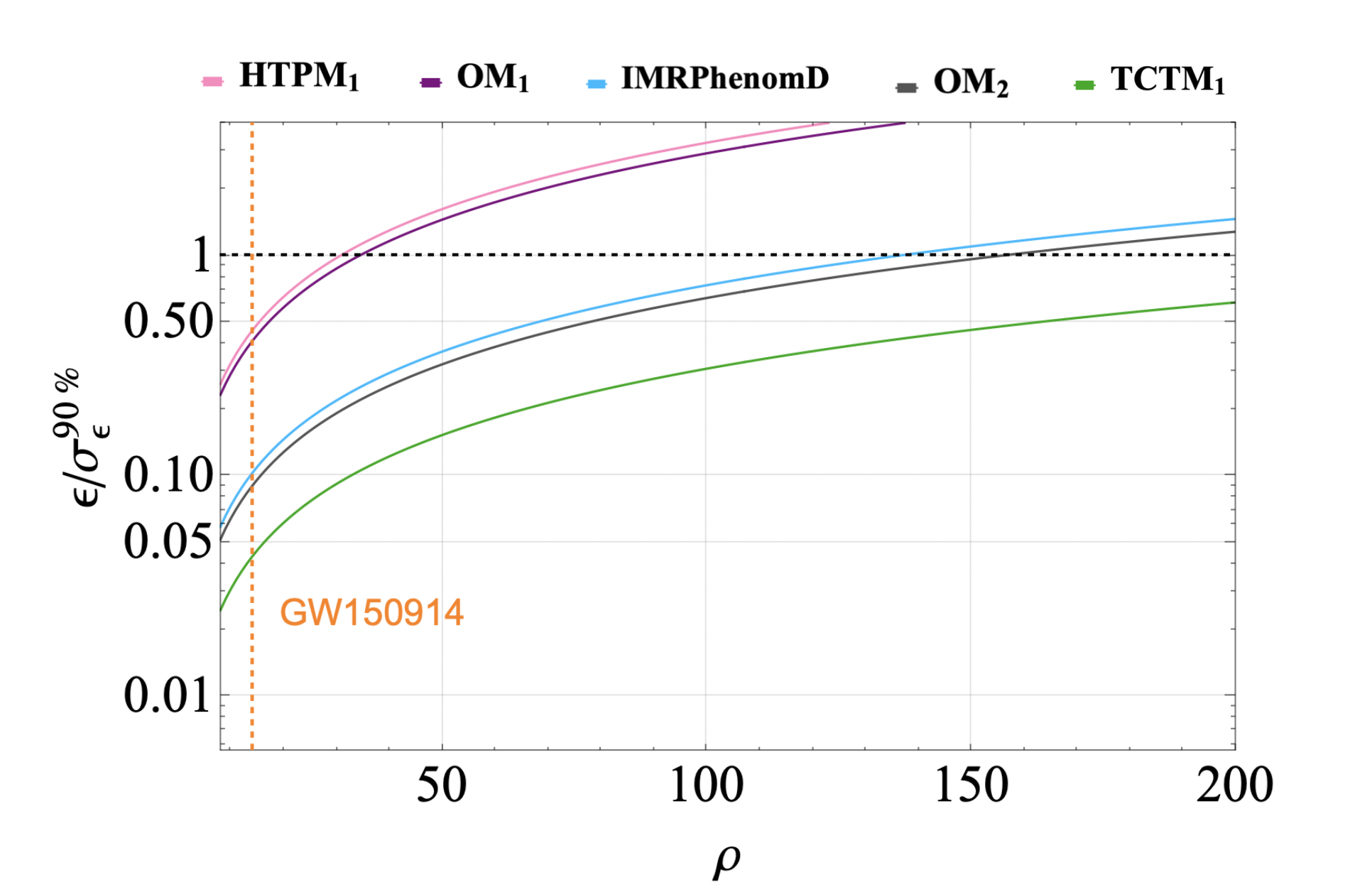}
\caption{The plot illustrates the ratio of the expected bias or systematic error $\epsilon$ to its $90\%$ credible interval statistical error $\sigma_\epsilon$ for each of the models examined in this study. The black dashed line, denoted by $\epsilon/\sigma_\epsilon \gtrsim 1$, signifies the point at which systematic error begins to dominate over statistical error. Notably, the models $\text{HTPM}_1$ and $\mathrm{OM}_1$ cross this threshold at RD $\rho\approx 30$, a region that may soon be reachable by current observatories or upcoming ones such as $A\#$. The orange dashed line corresponds to the RD $\rho$ value of the first event, GW150914.}
\label{fig:epsilon}
\end{figure}

In Fig.~\ref{fig:Mass_Spin_Comparison_model}, we show  the posterior distributions of the final mass and the final spin for the models $\text{OM}_2$, $\text{TCTM}_1$, $\text{HTPM}_1$ and IMRPhenomD for the same waveform at $\rho=100$. We compare the inferred posterior distributions of the optimal linear model $\text{OM}_2$ with all the other nonlinear models. Notice that the $90\%$ credible contours of the nonlinear IMRPhenomD model still provide tighter constraints on the remnant parameters and a lower value for for the bias $\epsilon$. It is noteworthy that IMRPhenomD is characterized by only four parameters, distinguishing itself from all the other models under consideration, each of which involves eight parameters. Another interesting observation is that even though $\text{TCTM}_1$ has better Bayes evidence compared to the corresponding $\text{OM}_2$ (i.e. having the same number of parameters), both its posterior distributions and its values for the $\epsilon$ are still compatible to each other. Notably, the model exhibiting the least favorable performance is $\text{HTPM}_1$, displaying a significant bias (approximately $0.2$) in relation to the true values. This bias arises from the model's flexibility in deviating from GR and from the mismodelling of the early time features by the GR-only $\mathrm{OM_1}$ model, resulting in a looser and biased constraint on the remnant mass and spin values. Significantly, the model $\text{HTPM}_1$ is presently employed for conducting GR tests in the analysis of ongoing events observed by the LVK collaboration thus, the use of this model might be limited to low $\rho$ scenarios, in which the statistical errors dominate the systematic ones. To this aim, in Fig.~\ref{fig:epsilon} we have quantified impact of the systematic errors with respect to the statistical ones. In particular, we have estimated the ratio between the systematic error $\epsilon$ and the statistical one $\sigma_\epsilon^{90\%}$ estimated at the $90\%$ percentile, $\epsilon/\sigma_{\epsilon}^{90\%}$, in terms of the SNR $\rho$. Notice that the quantity $\epsilon/\sigma_{\epsilon}^{90\%}$  helps one identify the SNR domain in which the systematic errors dominate, rather than providing any ranking based on the performance of the various models used in this work, since low values of this ratio favour large statistical uncertainties as much as low systematic errors. We note that the applicability of $\text{HPTM}_1$ and $\text{OM}_1$ appear to approach their limit around RD $\rho \approx 30$. At this point, the influence of systematic errors begins to overtake the statistical uncertainties in the inference of mass and spin values (combined in the variable $\epsilon$).
Yet, the prevailing the statistical uncertainty evident in present events, driven by the noise sensitivity of observatories, notably outweighs the parameter biases identified in this study. However, as demonstrated in our study, the accurate determination of the final mass and spin would become predominantly influenced by the systematic errors for the linear models currently used in ringdown parameter estimations (for LVK events) $\text{HTPM}_1$ and $\text{OM}_1$, when confronted with the anticipated high-$\rho$ scenarios characteristic of third-generation observatories. A much larger $\rho > 120$ is required for the simple non-linear models IMRPhenomD and $\text{TCTM}_1$, which show enough accuracy to address the 3G requirements at high $\rho$. An analogous analysis for the waveform Ext-CCE:BBH:0002 is presented in Appendix~\ref{app:extcce}.

\subsection{Dependence on the starting time $t_0$}
\label{sub:different-time}
Usually, there is a trade-off in truncating the waveform at different starting times. On the one hand, performing BH spectroscopy at late times RD leads to large statistical uncertainties due to the rapid decay of the SNR\footnote{The SNR $\rho$ scales inversely with the parameter uncertainty $\sigma_\lambda$ as $\rho\sim 1/\sigma_\lambda$. As a rule of thumb, $\rho(t_0 = 0)/\rho(t_0/M = 10) \sim e^{-1} = 0.37$, which implies that the posterior distributions on the physical parameters at $t_0/M =10$ are about $e$ times larger than the ones obtained at $t_0 = 0$.}. On the other hand, starting the spectroscopic analysis close to the peak amplitude of the strain, results in a biased estimation of the parameters. This bias stems from the absence of modes in the OMs and potentially ignoring the nonlinear effects~\cite{Cheung:2022rbm, Lagos:2022otp, Mitman:2022qdl}.

\begin{figure}[h]
\includegraphics[width=0.98\columnwidth]{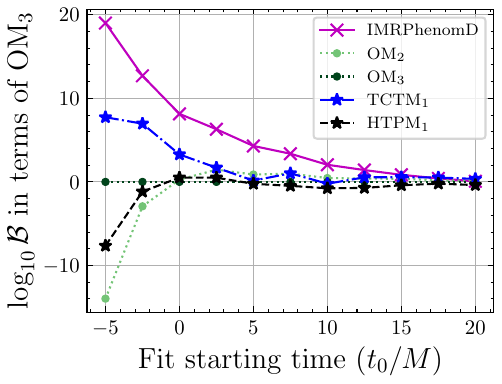}
    \caption{We show the evolution of $\log_{10}\mathcal{B}$ with respect to $\text{OM}_3$ for 4 different models, $\text{OM}_2$, $\text{TCTM}_1$, $\text{HTPM}_1$ and IMRPhenomD (as well as the reference $\mathrm{OM}_3$ at $\log_{10}\mathcal{B} = 0$ by construction), for a range of starting times $-5 \leq t_0/M \leq 20$, including times slightly before the amplitude peak ($t_0 / M = -5, -2.5$); at an assumed SNR $\rho=100$.
    }
    \label{fig:time-bayes}
\end{figure}

Here, we vary the fit starting time by truncating the waveform at different $t_0/M$, and we analyse the results in terms of the Bayes factor. We repeat the analysis for a set of starting times $t_0/M = -5, -2.5, 0, 2.5, \dots, 20$, where negative $t_0/M$ include part of the waveform slightly before merger. The runs have been performed for the models $\text{OM}_2$, $\text{OM}_3$, $\text{TCTM}_1$, $\text{HTPM}_1$ and IMRPhenomD with reference ($t_0 = 0$) SNR\footnote{%
Note that the actual SNR also relies on the choice of fit starting time $t_0$. Truncating and fitting the waveform from the peak can result in larger $\rho$ compared to fitting with only the later time of data. In order to make sure we consider the same event, we fix $\rho=100$ at $t_0=0$ and let the other $t_0\neq0$ SNR $\rho$ scaled with respect to the truncated waveform length.%
} $\rho_{t_0 = 0}=100$. In Fig.~\ref{fig:time-bayes} we show the $\log_{10}$ Bayes factors for each model with respect to the $\text{OM}_3$ model via Eq.~\eqref{eq:bayes factor}. We have found that the two nonlinear models considered in this work, IMRPhenomD and $\text{TCTM}_1$, provide positive log Bayes factors over $\text{OM}_3$, while the linear models $\text{OM}_2$ and $\text{HTPM}_1$ provide negative log Bayes factors around the merger (for negative and low positive $t_0/M$ values). In particular, the Bayes evidence values we have obtained strongly support the IMRPhenomD model over OMs at early starting times, including before the waveform amplitude peak\footnote{%
We note that both nonlinear models considered are strongly preferred over the linear ones ($\mathrm{OM}_2$, $\mathrm{OM}_3$, $\mathrm{HTPM}_1$) at times slightly before the peak, $t_0/M = -5, -2.5$. This is not very surprising for IMRPhenomD which is designed to consistently model pre- and post-merger phases, but it also holds, with lower evidences, for the nonlinear $\mathrm{TCTM}_1$.}%
, and up to $t_0/M=15$. This suggests that the nonlinear perturbations or even the prompt response effects could still dominate a short time after the merger. However, $\text{OM}_2$ consistently exhibits a comparable performance compared to all other models starting from $t_0=15$ onward, as the truncated waveform transitions into the linear regime.

\subsection{Can we observe the high-overtone amplitudes?}
 \begin{figure}[h]
\subfloat{
\includegraphics[width=0.98\columnwidth]{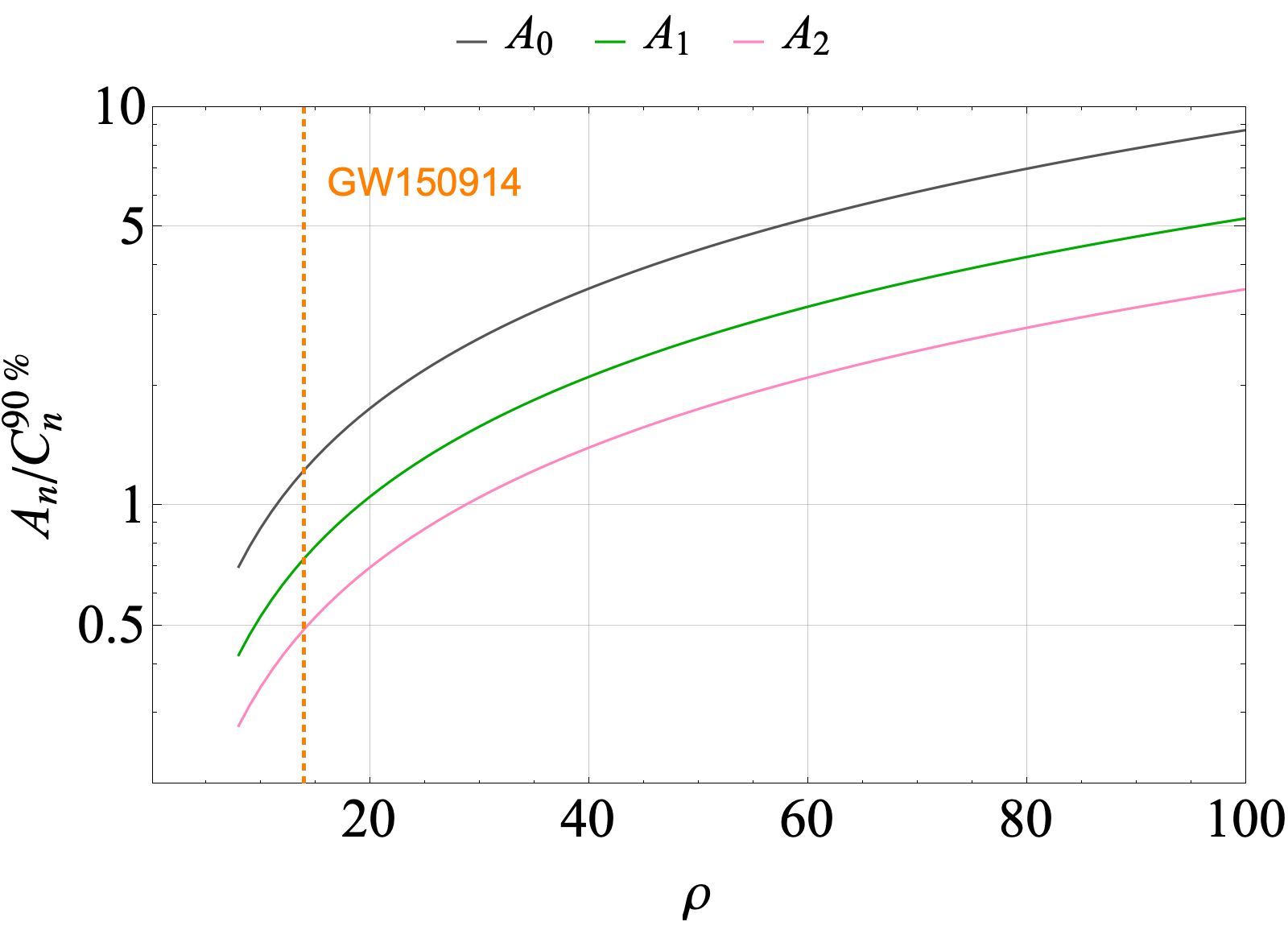}}\\
\subfloat{
\includegraphics[width=0.98\columnwidth]{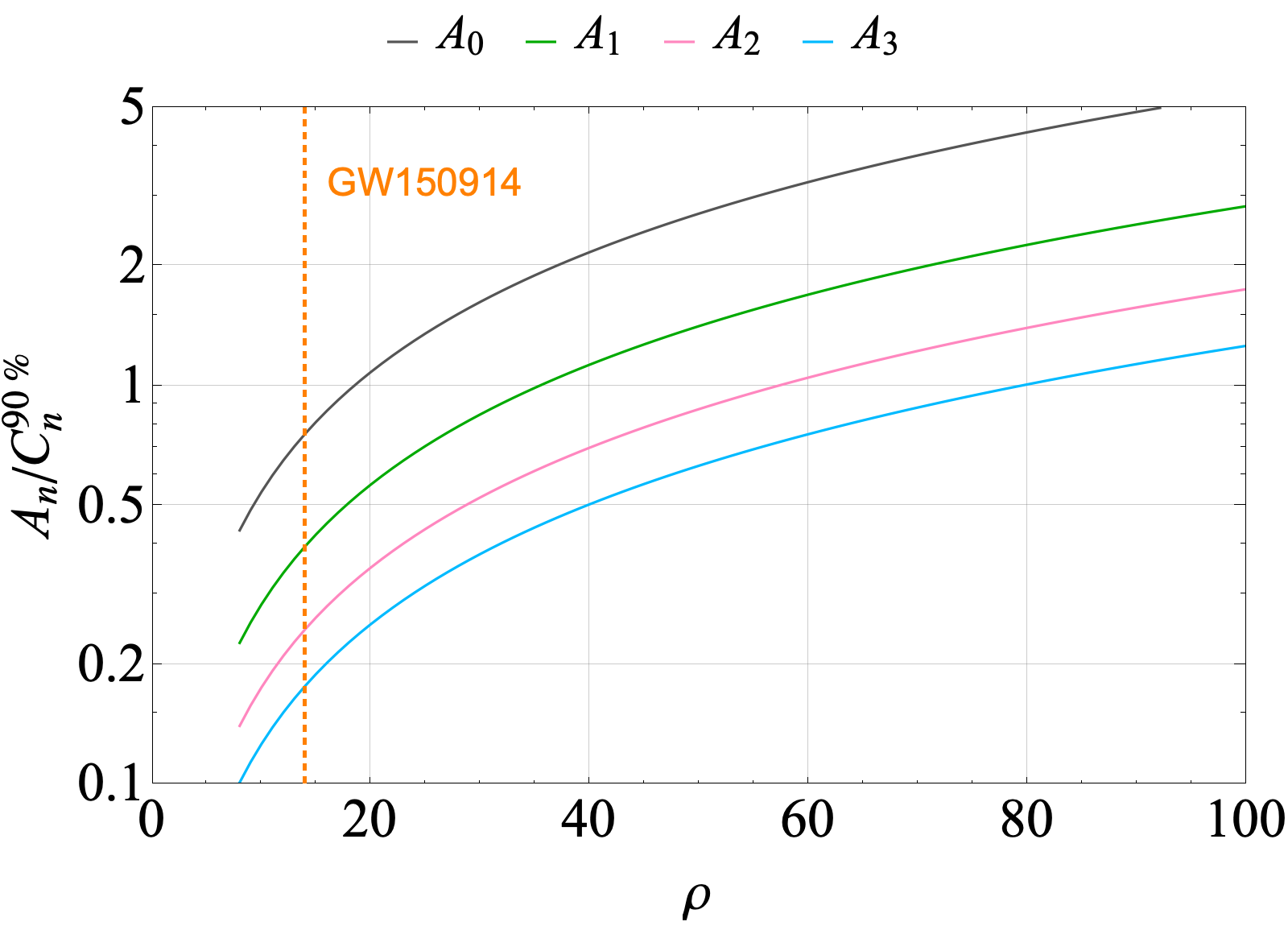}}
    \caption{Ratio $A_n/C_n^{90\%}$ of each tone's best-fit amplitude $A_n$ to its $90\%$ confidence interval, in terms of the simulated SNR for the GW150914-like waveform SXS:BBH:0305. The top panel represents $A_n/C_n^{90\%}$ for the $\mathrm{OM}_2$ \emph{i.e.} with $N_{\text{max}}=2$, and in the bottom panel we show the same results but for $N_{\text{max}}=3$. We take as a reference for observability $A_n/C_n^{90\%} \sim 1$. The vertical orange dashed line provides the post-peak SNR of GW150914, corresponding to a starting time $t_0 = 0$.
    }
     \label{fig:snr_amp}
\end{figure}

The confidence statistical observation of a tone solely from the RD regime, requires that the inferred value of its amplitude must be incompatible with zero, at least, within a given confidence level per tone amplitude $C_{n}$. Here, we use the one-sided $90\%$ confidence value $C_{n}^{90\%}$ to assert the confident observation\footnote{The posterior amplitudes, denoted as $A_n$  and obtained from running PE are typically constrained to be positive based on our chosen priors. This imposes a stringent requirement on the lower limit, specifically, that $C_{n}^{90\%}$ must be greater than zero. In this context, the one-sided $C_{n}^{90\%}$ is determined by assuming a Gaussian symmetry around the peak value, making it approximately equivalent to the upper bound of the $90\%$ confidence interval.} of a tone at a given SNR $\rho$ for the NR waveform SXS:BBH:0305~\cite{JimenezForteza:2020cve}. In particular,  we study the dependence of the magnitude $A_n/C_n^{90\%}$ for a set of SNRs injections with a network SNR $\rho \in \left[10,100\right]$, where $A_n$ provides the real valued amplitude of the tones. An approximated observation of a tone would correspond to $A_n/C_n^{90\%} \sim 1$. To estimate $C_{n}^{90\%}$ we use the framework described in Sec.~\ref{sec:setup}, where we also use the best fitting values of the amplitudes and phases obtained at $t_0$. In Fig.~\ref{fig:snr_amp} we show the quantity $A_n/C_n^{90\%}$ in terms of the SNR $\rho$ for two of the OMs considered in this work; $N_{\text{max}}=2$ (top panel) and $N_{\text{max}}=3$ (top panel). The dashed vertical orange line corresponds to the post-peak SNR of GW150914~\cite{isi2019,CalderonBustillo:2020rmh}. Notice that for each tone amplitude, the SNR required to cross one is larger as the $n$ index increases. This is because as $n$ increases the corresponding damping time of each tone $\tau_n$ decreases, thus carrying out a lower per-tone $\rho$. Specifically,  $\rho_{n=1}(A_0/C_0^{90\%}  =  1) > \rho_{n=1} (A_1/C_1^{90\%}= 1) > \rho_{n=2} (A_2/C_2^{90\%}=1) >\rho_{n=3} (A_3/C_3^{90\%} =1)$. Moreover, notice that $\rho \sim 30$ -- $A_2/C_2^{90\%}=1$ -- is  required for the full and confident observation of the $N_{\text{max}}=2$ model while $\rho \sim 80$ -- $A_3/C_3^{90\%}=1$-- is expected if $N_{\text{max}}=3$. A RD with $\rho \sim 30$ might be observed in the LIGO $A\#$ while for $\rho \sim 80$ is expected to occur only for 3G observatories as ET and CE~\cite{Asharp_sensitivity,Evans:2021gyd}. However, a significant concern with the high-order overtone models lies in the pronounced variation of tone amplitude when altering the number of tones, denoted as $N_{\text{max}}$ in the models~\cite{Bhagwat:2019dtm, Giesler:2019uxc}. For instance, the amplitude ratio between the amplitudes obtained from the fits to the NR data and for the ${N_{max}=2,3}$ models varies as 
$A_n^{N_{max}=3}/A_n^{N_{max}=2}= \lbrace 1.04, 1.50, 3\rbrace$ for the tones $n=0,1,2$ respectively, from the fits to the NR data. The substantial variability, on the order of $\mathcal{O}(2-3)$, in the amplitude values for tones $n=1,2$ poses a challenge in determining whether these values represent the system faithfully or are influenced by other fitting systematics, such as the absence of tones or the presence of nonlinearities. Conversely, the value of the fundamental tone amplitude $A_0$ remains almost constant regardless the number of tones considered in this example. These amplitude variation results are consistent with, \emph{e.g.}, \cite{Forteza:2021wfq,Baibhav:2023clw,CheungEtAl2023}.


\begin{table*}[!htb]
\begin{tabular}{C{0.1\textwidth}C{0.1\textwidth}C{0.06\textwidth}C{0.05\textwidth}C{0.05\textwidth}C{0.12\textwidth}C{0.1\textwidth}C{0.1\textwidth}C{0.1\textwidth}C{0.1\textwidth}}
\hline\hline
Catalog & Waveform & $q$ & $\chi_{1,z}$ & $\chi_{2,z}$ & $\text{IMRPhenomD}$ & $\text{OM}_2$ & $\text{OM}_3$ & $\text{TCTM}_1$ & $\text{HTPM}_1$  \\ \hline\hline
Main & BBH:0150 & 1    & 0.2      & 0.2  & (-8.215, 0.0080) & (-19.916, 0.0283) & (-22.449, 0.0095) & (-14.771, 0.0119) & (-19.314, 0.2059)  \\
& BBH:0305 & 1.221     & 0.33  & -0.44  & (-9.558, 0.0150) & (-17.032, 0.0270) & (-17.121, 0.0115) & (-14.447, 0.0246) & (-17.208, 0.2367) \\
& BBH:1221 & 3         & 0 & 0    & (-7.988, 0.0074) & (-16.652, 0.0494) & (-18.539, 0.0456) & (-15.071, 0.0503) & (-15.944, 0.2531)   \\
& BBH:0300 & 8.5       & 0   & 0  & (-6.207, 0.0092) & (-17.000, 0.1209) & (-18.583, 0.0851) & (-15.456, 0.0323) & (-15.647, 0.3011)  \\ 
\hline
Ext-CCE & BBH:0002 & 1 & 0.2 & 0.2 & (-44.57, 0.0238) & (-80.87, 0.0282) & (-81.44, 0.0238) & (-77.01, 0.0592) & (-79.67, 0.1880) \\
\hline\hline
\end{tabular}
\caption{
The table contains ($\log_{10}\mathcal{Z}$, $\epsilon$) calculated for models $\text{IMRPhenomD}$, $\text{OM}_2$, $\text{OM}_3$, $\text{TCTM}_1$ and $\text{HTPM}_1$, at $\rho=100$. They are sampled using the \textit{dynesty} Python package for five binary BH merger NR waveforms from both the main and Ext-CCE catalog in SXS. The properties of the initial BH binary in each case, i.e., the mass ratio and the two dimensionless spins are also given. Notice here that the $\text{OM}_2$ always provides a similar/slightly better evidence than $\text{OM}_3$, which is consistent with the results already obtained for SXS:BBH:0305.}
\label{table:bayes}
\end{table*}


\subsection{Analysis on different NR waveforms}
\label{sub:different_waveform}
To assess the robustness of our findings, we conduct parameter estimation using a distinct set of NR waveforms sourced from the SXS catalog. These waveforms, namely SXS:BBH:0150, SXS:BBH:0300, and SXS:BBH:1221, cover a spectrum of mass ratios spanning ${1, 3, 8.5}$ and effective spins of ${0.2, 0, 0}$. We also append waveform Ext-CCE:0002 from the Ext-CCE catalog to the list, with the detailed discussion presented in Appendix~\ref{app:extcce}. Notice that these waveforms, as well as SXS:BBH:0305 considered so far, were not used to calibrate IMRPhenomD~\cite{Khan:2015jqa}, thus are not biasing our model comparison. In Table~\ref{table:bayes} we show the $\log_{10}$ Bayes evidence and biases $\epsilon$ for a selected set of models, i.e., $\text{IMRPhenomD}$, $\text{OM}_2$, $\text{OM}_3$, $\text{TCTM}_1$ and $\text{HTPM}_1$ for all NR waveform truncated at a starting time $t_0=0$. 

We observe that IMRPhenomD consistently provides Bayes factors of approximately $\sim 10$ over OMs at $\rho=100$, which denotes a clear preference for the nonlinear IMRPhenomD model compared to linear RD models. In addition, the nonlinear model $\text{TCTM}_1$ emerges as the second best solution based on the Bayes evidences, showing a considerable advantage over the OMs. In particular, notice that we don't observe significant differences on the values of $\log_{10}\mathcal{Z}$ for the $\text{OM}_2$ and $\text{OM}_3$, indicating that $\text{OM}_2$ provides a sufficiently accurate solution at $t_0=0$, consistent with the results observed for SXS:BBH:0305. While the non-GR model $\text{HTPM}_1$ offers log evidence values similar to those of the OM models.

Regarding the bias analysis,  we have found that the IMRPhenomD model recovers the minimum value for $\epsilon$, for the waveforms SXS:BBH:0150, SXS:BBH:0300, and SXS:BBH:1221. $\text{TCTM}_1$ appears as the second best model with comparable/better epsilon values comparing to the $\text{OM}_2$ and $\text{OM}_3$. And $\text{HTPM}_1$ generally performs the poorest in recovering the true final parameters as expected, which is similar as the case we have seen for SXS:BBH:0305 analysis. Therefore, at $\rho=100$, for the five different NR waveforms tested here, we consistently observe the compelling preference for the nonlinear models compared to the OM models and non-GR model.

\section{Conclusions}
\label{sec:conclusion}
In this work we have tested the performance of several RD models by fitting them to the 22 mode of quasi-circular, non-precessing NR waveforms. The models we have used to fit the NR data are divided into four categories: i) The non-linear RD regime of the IMRPhenomD approximant --- which has been widely used in GW data analysis, ii) the RD model as described by linear QNMs --- dubbed here as OM,  iii) a family of non-linear RD toy models, the TCTMs, which expand upon the linear models by including further qualitative non-linear contributions, and iv) the HTPM models, which are linear but allow for deviations of the QNM spectrum from GR. Our results are obtained on NR waveforms of different nature: four extracted at finite radii and extrapolated to null infinity --- labelled as $\text{SXS:BBH:\#}$ --- and one extracted through the Cauchy characteristic procedure, thus, with lower expected errors that the extrapolated ones, and labelled as $\text{Ext-CCE:BBH:0002}$. We first analyse the performance of the models at fitting the $22$ mode of the waveform SXS:BBH:0305, at a starting time $t_0/M = 0$ (corresponding to the peak of amplitude of the $22$ mode). We obtain the minimum mismatch $\mathcal{M}$ and the bias $\epsilon$ on the recovered physical parameters of the remnant BH for all the models consideredand over a range of maximum number of tones included in the models. We observe that the non-linear TCTM models provide in general a lower mismatch than the overtone ones, even for $N_{\text{max}}=7$ --- while the IMRPhenomD post-peak model has a match to the data comparable to the 8-parameter $\nmax = 2$ OM model despite relying on only 4 free parameters. Regarding the final mass and spin recovery, we find that the non-linear models IMRPhenomD and TCTMs both show a similar accuracy to the OMs up to $N_{\text{max}}\sim 3$, while the OM solution outperforms them at $N_{\text{max}}\gtrsim 3$. Next, we have also performed injections of the NR waveform SXS:BBH:0305 in zero-realization white Gaussian noise at SNR $\rho = 100 $ and at different starting times $t_0/M\in[-5,20]$, simulating the expected SNRs of the third-generation gravitational wave observatories such as the Cosmic Explorer and Einstein Telescope. We first note that the non-linear model IMRPhenomD provides tighter constraints on the final mass and spin parameters at $t_0 = 0$ than the OM, even for $N_{\text{max}}=3$. In particular, $N_{\text{max}}=2$ and $3$ provide compatible but broader mass and spin contours with IMRPhenomD albeit with a significantly lower Bayes evidence, while the $\nmax=0,1$ OM models are more biased, which is consistent with the results shown in~\cite{CalderonBustillo:2020rmh}. Remarkably, there is currently no amplitude and phase overtone model calibrated to the BBH progenitor masses and spins for $N_{\text{max}}>1$~\cite{london:2014cma,JimenezForteza:2020cve,CheungEtAl2023}, which contributes to the broader $\nmax=2,3$ OM constraints. If such a model were to be developed, the additional uncertainty introduced by the calibration process might be expected to further broaden the amplitude and phase uncertainties, potentially compromising the accuracy compared to the IMRPhenomD model.

Moreover, we have computed the $\log_{10}$ Bayes factor of the OM, HTPM, IMRPhenomD and TCTM models with respect to ${\text{OM}_3}$ (\emph{i.e.}, OM with $N_{\text{max}}=3$). We have found that in terms of the Bayes factor, at $t_0=0$, the IMRPhenomD is the preferred model, the second best ranked model being the $\text{TCTM}_1$: $\log_{10} \mathcal{B}^{\text{IMRPhenomD}}_{\text{OM}_3}\sim 8$ and $\log_{10} \mathcal{B}^{\text{IMRPhenomD}}_{\text{TCTM}_1} \sim 5$, showing a decisive (in the vocabulary of~\cite{Kass:1995loi}) evidence towards the IMRPhenomD model. The preference for the nonlinear IMRPhenomD model remains consistent until $t_0/M \sim 15$, at which point the ${\text{OM}_2}$ exhibits a comparable evidence. It is also important to note that the non-linear $\text{TCTM}_1$ model consistently offers superior fits to the data compared to the OM models at negative/early starting times. However, as anticipated, this difference also diminishes at late fitting starting times. Next, we have estimated the SNR required for observing confidently a tone amplitude for the OM models, with $N_\text{max} = 1, 2, 3$. First, we have obtained that, at $90\%$ credible level, the ratio of bias to statistical uncertainty $\epsilon/\sigma_\epsilon^{90\%}$ exceeds one for the models $\text{OM}_1$ and $\text{HTPM}_1$ when $\rho \sim 30$. This implies that analyses of the RD of upcoming loud GW events might be soon dominated by the systematic errors if those models are used. For the nonlinear and the $\text{OM}_{n \geq 2}$ models, we show that the systematic errors remain subdominant up to $\rho \sim 150$. Additionally, we observe that achieving a full observation and/or characterization of all the amplitude parameters of the linear models $\text{OM}_{n=2,3}$ would require a SNR $\rho \sim 30, 80$ respectively. However, the strong variability of their amplitudes pose reasonable doubts on the physical reliability of the models as compared to the full nonlinear solutions, which simply depend on the well-known progenitor parameters. Finally, we carried out parameter estimation with the same set of models on four additional waveforms from the SXS catalog, including one example from its Ext-CCE extension. We observe again the compelling preference for the nonlinear models compared to the OM and non-GR (HTPM) models regarding the evidence as well as (in the case of the IMRPhenomD model) the recovery bias, showing the robustness of the trends observed in our more detailed SXS:BBH:0305 analysis.

In summary, we have performed ringdown PE on five independent quasi-circular NR simulations (not part of the calibration set of the IMR model considered) over a range of mass ratios and aligned, anti-aligned or vanishing spins. Our findings indicate that IMR-based nonlinear models such as the IMRPhenomD model, yield higher Bayes factors than the QNM-only models, especially in high-SNR ($\rho\approx100$) scenarios. This implies a higher accuracy than QNM models in fitting NR ringdown waveforms up to early times. Moreover, the IMRPhenomD model results in tighter posterior distributions for the black hole final mass and spin. Additionally, we observe that the value of the evidence of the OM models saturates at $\nmax = 2$. The decrease of the evidence at $N_{\text{max}}>2$ could be induced by the expected non-linearities affecting the early post-peak phase which, at the same time, may be causing the observed large instabilities on the amplitudes of the tones with $n>1$~\cite{Bhagwat:2019dtm,Forteza:2021wfq,Baibhav:2023clw,Cheung:2023vki,Zhu:2023mzv}. For our toy nonlinear model $\text{TCTM}_1$, we observe similar posterior distributions than for the OM models, together with a better Bayes factor at early times --- intermediate between that of OM models and that of the IMRPhenomD model. Tests with other NR calibrated models as the IMRPhenomPv2 and SEOBNRv4 approximants~\cite{hannam:2013oca,t1500602,bohe:2016gbl} provided the same qualitative results. 

Hence, we conclude that the utilization of nonlinear models and, especially, well-established IMR models which are calibrated to the progenitor parameters, can be more pertinent when inferring physical parameters from a non-precessing, quasi-circular RD signal. This is particularly relevant when the analysis commences at the peak of the strain and is applicable to SNR ratios consistent with third-generation observatories such as ET and CE. Further investigation into other NR waveforms from different catalogues, specifically focusing on precessing binaries, will be performed elsewhere.

\begin{acknowledgements}
This paper was initiated as a result of an internship carried out virtually under the Max Planck Institute for Gravitational Physics Hanover. We acknowledge the Max Planck Gesellschaft for support and we are grateful to the Atlas cluster computing team at MPG Hanover for their help. The authors are also thankful to Alex~Nitz, Collin~Capano, Sumit~Kumar, and Yifan~Wang for useful discussions, to Juan~Calderon~Bustillo, Gregorio~Carullo, Will~Farr and Vasco~Gennari for helpful comments on the draft, and to an anonymous referee for insightful comments. P.~M. and X.~J.~F. are supported by the Universitat de les Illes Balears (UIB); the Spanish Agencia Estatal de Investigaci\'on Grants No. PID2022-138626NB-I00, No. RED2022-134204-E, and No. RED2022-134411-T, funded by MCIN/AEI/10.13039/501100011033/FEDER, UE; the MCIN with funding from the European Union NextGenerationEU/PRTR (No. PRTR-C17.I1); the Comunitat Auton\`oma de les Illes Balears through the Direcci\'o General de Recerca, Innovaci\'o I Transformaci\'o Digital with funds from the Tourist Stay Tax Law (No. PDR2020/11 - ITS2017-006), and the Conselleria d'Economia, Hisenda i Innovaci\'o Grant No. SINCO2022/6719. X. J. F. is also supported by the Spanish Ministerio de Ciencia, Innovación y Universidades (Beatriz Galindo, BG22-00034) and cofinanced by UIB.
\end{acknowledgements}

\bibliography{draft.bib}
\appendix

\section{Robustness tests of prior choices}
\label{app:prior}

\begin{table}[h]
\begin{tabular}{C{0.15\textwidth}C{0.15\textwidth}C{0.15\textwidth}}
\hline\hline
Model & Parameter & Prior \\ \hline\hline
$\text{OM}_{0}$ & Remnant mass    & (0.5, 1.3)   \\
 & Remnant spin    & (0, 0.99)   \\
 & Amplitudes    & (0, 2)   \\
 & Phases    & (0, 2$\pi$)   \\ \hline
$\text{OM}_{1\sim2}$ & Remnant mass    & (0.5, 1.3)   \\
 & Remnant spin    & (0, 0.99)   \\
 & Amplitudes    & (0, 10)   \\
 & Phases    & (0, 2$\pi$)   \\ \hline
$\text{OM}_{3\sim4}$ & Remnant mass    & (0.5, 1.1)   \\
 & Remnant spin    & (0, 0.99)   \\
 & Amplitudes    & (0, 10)   \\
 & Phases    & (0, 2$\pi$)   \\ \hline
$\text{TCTM}_{n}$ & Remnant mass    & (0.5, 1.3)   \\
 & Remnant spin    & (0, 0.99)   \\
 & Amplitudes    & (0, 2)   \\
 & Phases    & (0, 2$\pi$)   \\ 
 & $\log_{10} A$    & (-5, 5)   \\
 & $\tau$    & (0, 100)   \\ \hline
 $\text{HTPM}_{n}$ & Remnant mass    & (0.5, 1.3)   \\
 & Remnant spin    & (0, 0.99)   \\
 & Amplitudes    & (0, 2)   \\
 & Phases    & (0, 2$\pi$)   \\ 
 & $\alpha$    & (-0.5, 1)   \\
 & $\beta$    & (-0.5, 1)   \\ \hline
IMRPhenomD & Mass ratio    & (1,8)   \\
 & Initial spin 1    & (-0.99, 0.99)   \\
 & Initial spin 2    & (-0.99, 0.99)   \\
 & Phases    & (0, 2$\pi$)   \\ 
\hline\hline
\end{tabular}
\caption{
The table provides an overview of the priors used for all the models we consider in the paper. Note that we use a log-flat prior for $A$ in $\text{TCTM}$ as its order of magnitude is not known \emph{a priori}; while all other parameters are uniformly distributed. Note that we use different priors for the amplitudes in different OM. The reasoning behind is that we observe broader, more loosely constrained posteriors in the higher overtones' amplitudes, hence the prior ranges need to be extended accordingly. While for the remnant mass, a tighter choice of priors on higher overtone models will rule out the local minimum and thus represent a more physical analysis on OM.}
\label{table:priors}
\end{table}

In parameter estimation, the choice of prior range is always a crucial factor. In fact, even if the posterior distributions are rather similar, the Bayes evidence can still be different for different prior assumptions. As shown in Table~\ref{table:priors}, in our case, since we want to compare between the IMRPhenomD and OMs, the priors already differ in the first place that one is a IMR model that requires initial properties of the binary, while the another is only a RD model. For IMRPhenomD, the mass ratio of (1, 8) prior range + (-0.99, 0.99) spins priors would lead to a different prior range on the remnant mass after conversion. The new prior range is $\sim$ (0.85, 1) for the final mass, which is much different from the typical prior range we use for overtone model, i.e. (0.5, 1.3).

\begin{figure}
\includegraphics[width=1\columnwidth]{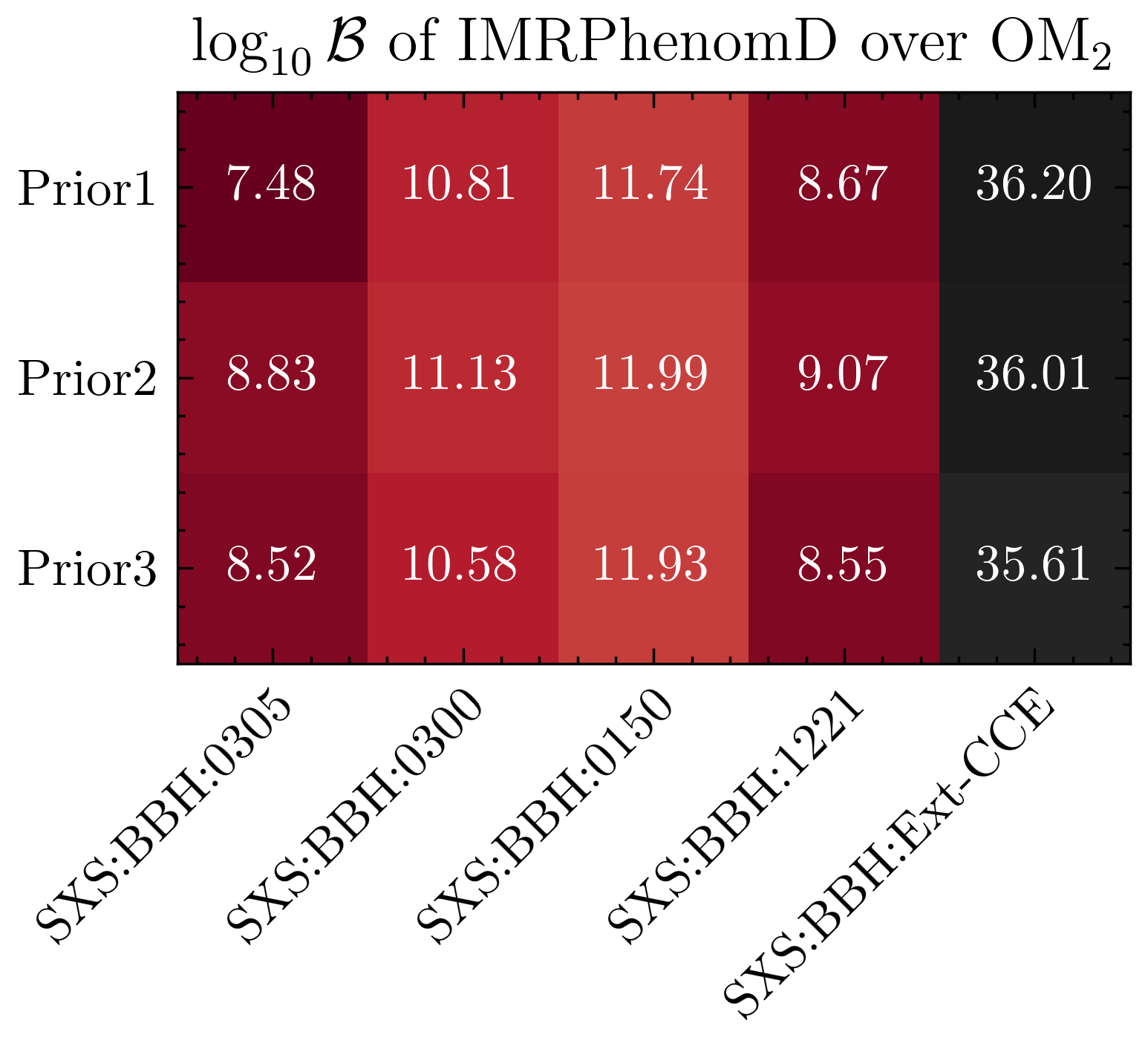}
    \caption{We show the $\log_{10}\mathcal{B}$ between IMRPhenomD and $\text{OM}_2$ for 3 different prior ranges and for 5 different waveform. In this plot, the "Prior1", "Prior2", "Prior3" are used to denote different ranges for the final mass prior of, (0.5, 1.3), (0.7, 1.2) and (0.8, 1.1), respectively. To examine the robustness of the choice of prior ranges, we compare the vertical 3 values in each column of the plot as they are computed on samplings of different waveform. Comparing the values within a column and for the three columns of the plot, we do not observe significant differences. This suggests that the choices made on the priors, do not significantly influence the relative performance of the RD models.
    }
    \label{fig:prior_test}
\end{figure}

To examine the robustness of our tests on different prior assumptions, we also perform the same sampling for $\text{OM}_2$ with 2 new different priors on particularly the final mass. They are (0.7, 1.2) and (0.8, 1.1), respectively. In Fig.~\ref{fig:prior_test}, the two new priors are denoted as "Prior2" and "Prior3" with the "Prior1" being (0.5, 1.3). In this plot, we show the $\log_{10}\mathcal{B}$ between IMRPhenomD and $\text{OM}_2$ for 3 different prior ranges and for 5 different waveform. To examine the robustness of the choice of prior ranges, we compare the vertical 3 values in each column of the plot as they are computed on samplings of the same waveform. We find that the largest difference in the $\log_{10}$ Bayes evidence we observe is in SXS:BBH:0305, which has a $\log_{10}\mathcal{B}\sim 1$ for one of the new prior sampling comparing to the typical prior sampling. However, this change is rather little compared to the absolute values of the Bayes factors, which have $\log_{10}\mathcal{B}>7$ for all the priors. Moreover, we don't see any clear trend in the improvement/reduction of Bayes factors when the prior range shrinks, i.e., from "Prior1" to "Prior3". Therefore, the difference we observe are also possibly subject to the statistical errors of the samplings. Not to mention in other tests, the difference between the Bayes evidence are even smaller. Therefore, we conclude that the prior choice of our test is robust.

\section{Posteriors of Ext-CCE waveform}
\label{app:extcce}
As another cross-check, we also present our bias analysis for the Cauchy characteristic Ext-CCE waveform from the additional SXS catalog particularly in terms of the mass/spin posterior distribution comparison of different models. The specific waveform we test is Ext-CCE:BBH:0002, which has true final dimensionless mass and spin $m_f$, $a_f$ given as [0.946, 0.746] 

\begin{figure}
\includegraphics[width=0.98\columnwidth]{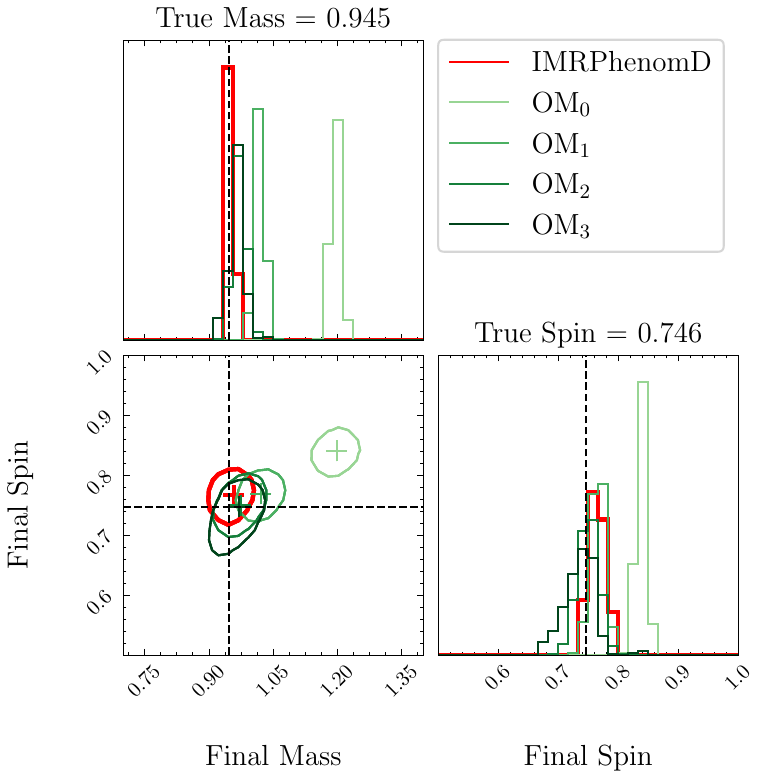}
    \caption{We show the comparison of final mass and spin posterior distribution for the sampling of $\text{OM}_{0-3}$ and IMRPhenomD for waveform Ext-CCE:BBH:0002 as a cross check. Each contour represents a $90\%$ credible region on the mass-spin 2D plane for given model. The black dashed lines note the "true" final mass and spin. The "+" symbols denote the maximum likelihood values for each of the models considered. We assume the injection $\rho$ to be 100 for all the cases, in order to see the clear comparison.
    }
    \label{fig:Mass_Spin_Comparison_OM_Ext}
\end{figure}

\begin{figure}
\includegraphics[width=0.98\columnwidth]{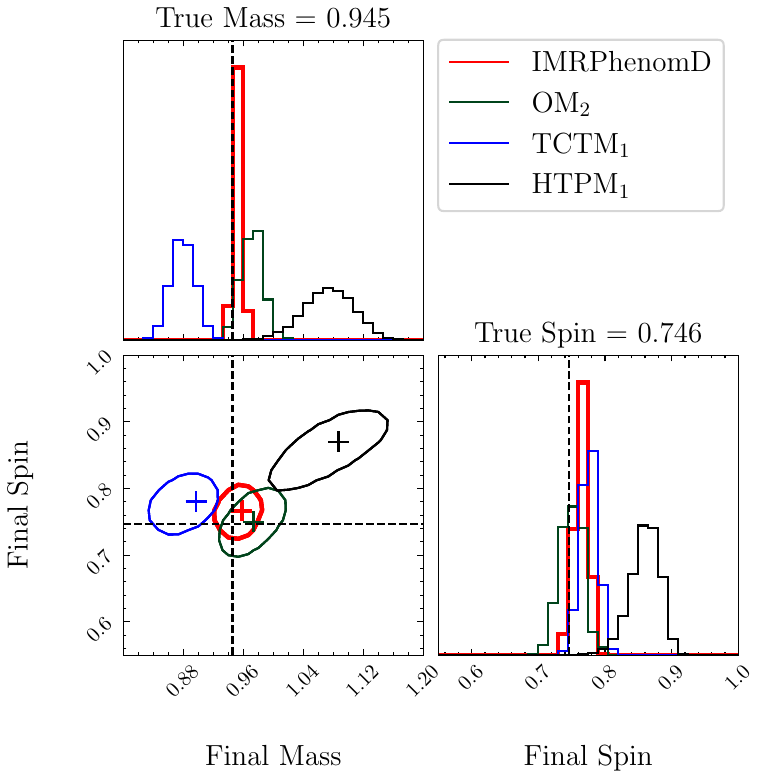}
    \caption{We show the comparison of final mass and spin posterior distribution for the sampling of $\text{OM}_2$, $\text{TCTM}_1$, $\text{HTPM}_1$ and IMRPhenomD for waveform Ext-CCE:BBH:0002. Each contour represents a $90\%$ credible region on the mass-spin 2D plane for given model. The "+" symbols denote the maximum likelihood values for each of the models considered here. The black dashed lines note the "true" final mass and spin. 
    }
    \label{fig:Mass_Spin_Comparison_model_Ext}
\end{figure}

In Fig.~\ref{fig:Mass_Spin_Comparison_OM_Ext}, we show firstly the comparison of final mass and spin posterior distribution for the sampling of $\text{OM}_{0-3}$ and IMRPhenomD. In the main text, we see that the posterior distributions improvement in successive OMs. Similarly, we also see here that the posterior distributions on the remnant mass and spin for OM improves significantly from $\text{OM}_0$ to $\text{OM}_1$, but then converges at $\text{OM}_3$. In particular, the epsilons values estimated for $\text{OM}_2$ and $\text{OM}_3$ are comparable, as the maximum likelihood values for both models almost overlap. The IMRPhenomD model here also shows a comparable parameter estimation performance to the $\text{OM}_{2,3}$ models. IMRPhenomD gives a more faithful recovery for the final mass, while it performs slightly worse in terms of the final spin as we can see from the 1D marginalized distribution. Therefore, the results obtained for the Ext-CCE:BBH:0002 waveform is consistent with the results obtained for SXS:BBH:0305 presented in the main text.

In Fig.~\ref{fig:Mass_Spin_Comparison_model_Ext}, we show then the comparison of final mass and spin posterior distribution for the sampling of $\text{OM}_2$, $\text{TCTM}_1$, $\text{HTPM}_1$ and IMRPhenomD assuming $\rho=100$. Again, this is test about different types of models. We can see from the $90\%$ credible contours that the nonlinear IMRPhenomD model performs the best in both the recovery accuracy and precision with only 4 free parameters. While for all the other models we have 8 parameters. While unlike the conclusion in the main text, here the another nonlinear model, $\text{TCTM}_1$ has much worse posterior distributions comparing to the corresponding $\text{OM}_2$ (i.e. having the same number of parameters). The model with the worst performance here is still $\text{HTPM}_1$ as it considers the deviation to GR, which leads then to a looser/biased constraint on the remnant mass and spin. 

\end{document}